\documentclass[sn-apa]{sn-jnl}

\usepackage{float}
\usepackage{siunitx}
\usepackage{textcomp}
\usepackage{amsmath,amssymb}
\usepackage{hyperref}

\begin{document}

\title[Shape perception integrates intuitive physics and analysis-by-synthesis]{3D Shape Perception Integrates Intuitive Physics and Analysis-by-Synthesis}

\author*[1,2,3]{Ilker Yildirim}\email{ilker.yildirim@yale.edu}
\equalcont{These authors contributed equally to this work.}
\author*[4,5]{Max H. Siegel}\email{maxs@mit.edu}
\equalcont{These authors contributed equally to this work.}
\author[4,5,6]{Amir A. Soltani}
\equalcont{These authors contributed equally to this work.}
\author[7]{Shraman Ray Chaudhari}
\author*[4,5]{Joshua B. Tenenbaum}\email{jbt@mit.edu}

\affil*[1]{Department of Psychology, Yale University, New Haven, CT}
\affil[2]{Department of Statistics \& Data Science, Yale University, New Haven, CT}
\affil[c]{Wu-Tsai Institute, Yale University, New Haven, CT}
\affil[4]{Department of Brain \& Cognitive Sciences, MIT, Cambridge, MA}
\affil[5]{The Center for Brains, Minds, and Machines, MIT, Cambridge, MA}
\affil[6]{Department of Psychology, Boston College, Chestnut Hill, Massachusetts, USA}
\affil[7]{Department of Electrical Engineering \& Computer Science, MIT, Cambridge, MA}

\abstract{

Many surface cues support three-dimensional shape perception, but people can sometimes still see shape when these features are missing -- in extreme cases, even when an object is completely occluded, as when covered with a draped cloth. We propose a framework for 3D shape perception that explains perception in both typical and atypical cases as analysis-by-synthesis, or inference in a generative model of image formation: the model integrates intuitive physics to explain how shape can be inferred from deformations it causes to other objects, as in cloth-draping. Behavioral and computational studies comparing this account with several alternatives show that it best matches human observers in both accuracy and response times, and is the only model that correlates significantly with human performance on difficult discriminations. Our results suggest that bottom-up deep neural network models are not fully adequate accounts of human shape perception, and point to how machine vision systems might achieve more human-like robustness.

}

\maketitle

\section*{Introduction}

\begin{figure*}[t!]
    \centering
    \includegraphics[width=.7\linewidth]{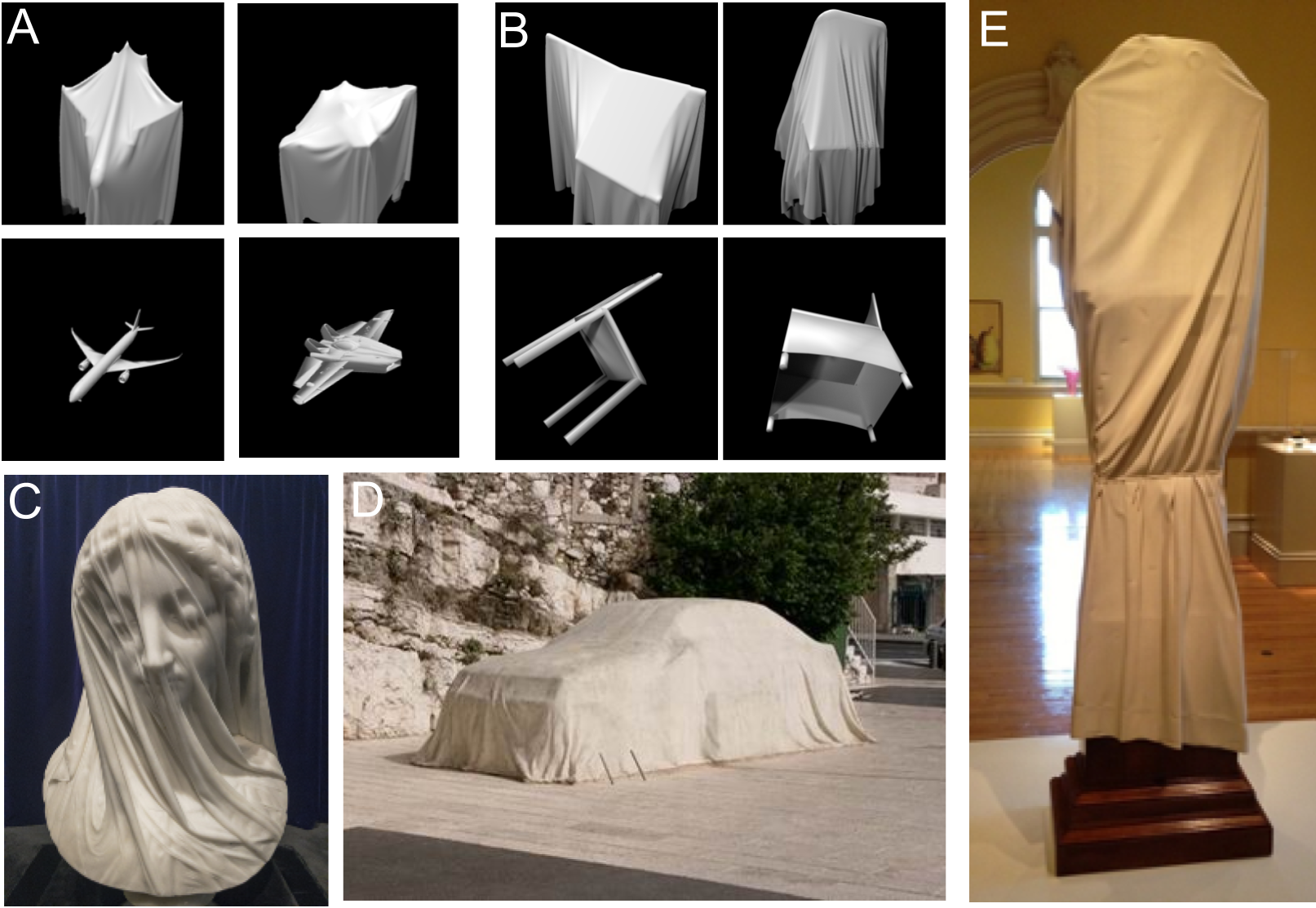}
    \caption{Seeing 3D shape through a cloth.
      (A) Bottom row: two different airplanes. Top row: the same airplanes, draped with cloth and presented in random order and in a random pose. (B) Same as  (A) but for two different chairs.
      Despite the variation in viewpoint and complete occlusion of the cloth-draped objects,  human observers can still match the cloth-draped and unoccluded object pairs. (See note in main text for correct answer.) (C-E) Sculptors have long displayed their skill in works that are crafted from a single rigid material but convey both an illusory effect of cloth draping and rich 3D shape for the object under the illusory cloth, as in Giovanni Strazza's (c. 1850s) ``Veiled Virgin'' (C, marble), Gabriel Klasmer's (2000) ``Car in the Sun'' (D, fiberglass), or Wendell Castle's (1985) ``Ghost Clock'' (E, mahogany, partially bleached).}
    \label{fig:intro-demo}
\end{figure*}

For more than a century, vision scientists have studied the many cues that humans or machines use to recover shape. Edges or bounding contours, gradients of shading or texture, stereo disparity, and motion parallax are just a few of the cues that can be computed from the visible surface of an object and that can reliably indicate an object's three-dimensional (3D) shape across different views \citep{bulthoff1991shape}. When available, surface cues effectively support shape perception in humans and machines. %
However, a set of recent studies \citep{yildirim2016perceiving,phillips2020veiled, little2021physically,wong2022seeing} present a challenge to the classical cue-based theory of shape perception:
Even when a surface is obscured, humans can sometimes perceive shape, without directly seeing the object %
at all.
Consider the synthesized images of cloth-covered objects in Fig. \ref{fig:intro-demo}A and B, in which each object is completely occluded by a thin, cotton-like fabric draping over it. Although the draped shapes look very different from the comparison objects (shown in randomly chosen orientations), observers can nonetheless pick out which unoccluded airplane or chair matches the 3D shape of the appropriate occluded object.  %
\footnotemark{}

Here we ask: How can people perceive object shape (and pose, size, category, etc.) in these images, when all the classic visual cues are mostly or entirely absent? Even those image cues that are present may be highly misleading, as they belong not to the underlying object's shape but the surface of the occluding cloth.
Somehow we are able to interpret the shape of the cloth as an interaction between the underlying rigid object's geometry and the way the cloth drapes and deforms upon contact. While sculptors have long exploited this capacity of the visual system to depict human faces and figures,  only recently have detailed behavioral studies provided convincing evidence that humans somehow ``undo'' \citep{phillips2020veiled, wong2022seeing} the effect of the cloth to access the hidden object. Thus far, a computational account of vision that can explain shape perception, even in the absence of surface cues, remains absent.

\footnotetext{The correct shape matches in Fig. 1A and B are: (A) left airplane on top goes with left on bottom, right on top goes with right on bottom; (B) left chair on top goes with right on bottom, right on top goes with left on bottom.}

There are at least two ways one might explain how people perceive the shape of draped objects, corresponding to two contemporary %
frameworks which each seek to advance beyond classical cue-based approaches to 3D shape. One possibility %
is that instead of a relatively constrained
set of interpretable, meaningful cues, often derived from an analysis of the geometry of objects and of image formation processes, the human visual system might engage a much larger set of cues which are obtained through some black-box learning mechanism (and are therefore difficult %
to write down or interpret). Recent computer vision models based on deep convolutional neural networks (DCNNs) have demonstrated learned feature hierarchies which facilitate impressive object recognition capabilities \citep{lecun2015deep,krizhevsky2012imagenet} and which are relatively robust to variation in appearance and pose even though the model training objective does not explicitly include these goals. Moreover, these same features have been shown to enable many other seemingly disparate visual tasks, including shape perception, with only minor adaptation (e.g. fine-tuning, or adding one or a small number of additional output layers) \citep{hong2016explicit}. Perhaps these features are sufficiently robust to generalize across even more extreme image transformations, such as cloth occlusion.

A second possibility is that we see 3D shape via ``analysis by synthesis'', or inference in a physics-based generative model of how scenes form and give rise to images \citep{yuille2006vision,mumford1994neuronal}. On this view, shape perception is not driven solely or primarily by a fixed, universal set of image cues, computed bottom-up from any image and sufficient for any downstream task. Rather, we infer 3D shape through a top-down interpretation process based on an internal model of how images are formed and the role that shape plays in that model.
The generative model approach sees cloth draping as just one exemplar of a potentially unbounded space of atypical presentations of objects, in which some aspects of the physics of scenes and images grossly alter an object's appearance from its typical form while remaining easily interpretable by humans: consider as other examples viewing an object such as the chair or airplane in Fig.~\ref{fig:intro-demo} outside in a rainstorm, or under ten feet of cloudy water, or through colored plastic wrap, or in the light of a full moon at night. The open-ended compositionality of the visual world may imply that it is difficult or impossible to specify or learn a single set of bottom-up image cues or features which reliably and robustly encode an object's 3D shape even in such atypical, rare conditions.
Instead, a system should model individual, scene-level causes -- the physical objects and processes that generate images -- and how they combine and yield visual input. Then, by reversing their effects, it may recover the original physical scene. Thus a visual system could still identify a draped object and even perceive its fine-grained 3D shape if it were able to model and somehow invert cloth physics.

Our goal in this paper is to use the perception of objects under cloth as a case study to evaluate concrete versions of each of these accounts of shape perception. The theoretical merits of the pure bottom-up and top-down approaches have been extensively debated in the literature, but it has been difficult to find strong evidence distinguishing the bottom-up cue-based and top-down model-based approaches; until recently, neither discriminative classifiers nor Bayesian generative models performed well in realistic visual tasks, so comparisons with biological vision were limited to controlled scenarios with simplified, non-naturalistic stimuli \citep{liu1995object}.  Advances in algorithms and computing hardware, however, have led to DCNN and analysis-by-synthesis models that achieve good performance with complex natural images and can now be rigorously evaluated as models of how we perceive 3D shape in challenging cases with rich naturalistic stimuli. They also let us explore various hybrid accounts that to date have received very little direct evaluation in human psychophysics: in particular, we compare human judgments with top-down analysis-by-synthesis models attempting to match images at either the level of raw pixels or intermediate-level representations based on DCNN features.

To our knowledge, only one empirical study has studied human perception in light of these improved methods. \cite{erdogan2017visual} defined a compositional generative model of 3D shape and compared human judgments of shape similarity\footnote{\cite{erdogan2017visual} elicited graded similarity judgments between different objects %
(so that the study and target items could be similar but not identical).
In contrast, we use forced-choice judgments of same or different object identity; the correct shape is always a response option.} with those derived from bottom-up classifiers and from top-down inference in a 3D generative model, %
concluding that the latter might underlie shape perception because it correlated better with human responses. But this study, while pioneering, provides only limited evidence for top-down analysis-by-synthesis in perception. The best bottom-up model was also quantitatively predictive of human judgments and performed almost as well as the top-down model. %
With more training and improved DCNNs, the quantitative gap between these models might be expected to narrow even further. In addition, neither model improved dramatically over a simple baseline matching test to target images at the pixel level. Here, we demonstrate a stronger, qualitative distinction between model classes enabled by our completely occluded cloth-draped stimuli: standard DCNNs and pixel-based observers, unlike people and our generative models, perform no better than chance on harder instances and even with extensive specialized training show little improvement.

Our design choices offer several other advantages. Because we chose uncommon stimuli with variable difficulty, we find meaningful variance in human performance and response time, which allows for finer-grained model evaluations and comparisons of humans and models on trial-by-trial speed as well as accuracy. The generative model that we consider performs iterative inference with significant stimulus-driven variation in the number of computational steps and therefore can be directly compared with subjects' reaction times, potentially revealing signatures and roles for feedback or recurrence in biological shape perception. %

\section*{Results}
\subsection*{The Object-Under-Cloth Task}

\begin{figure*}[ht!]
    \centering
    \includegraphics[width=0.9\linewidth]{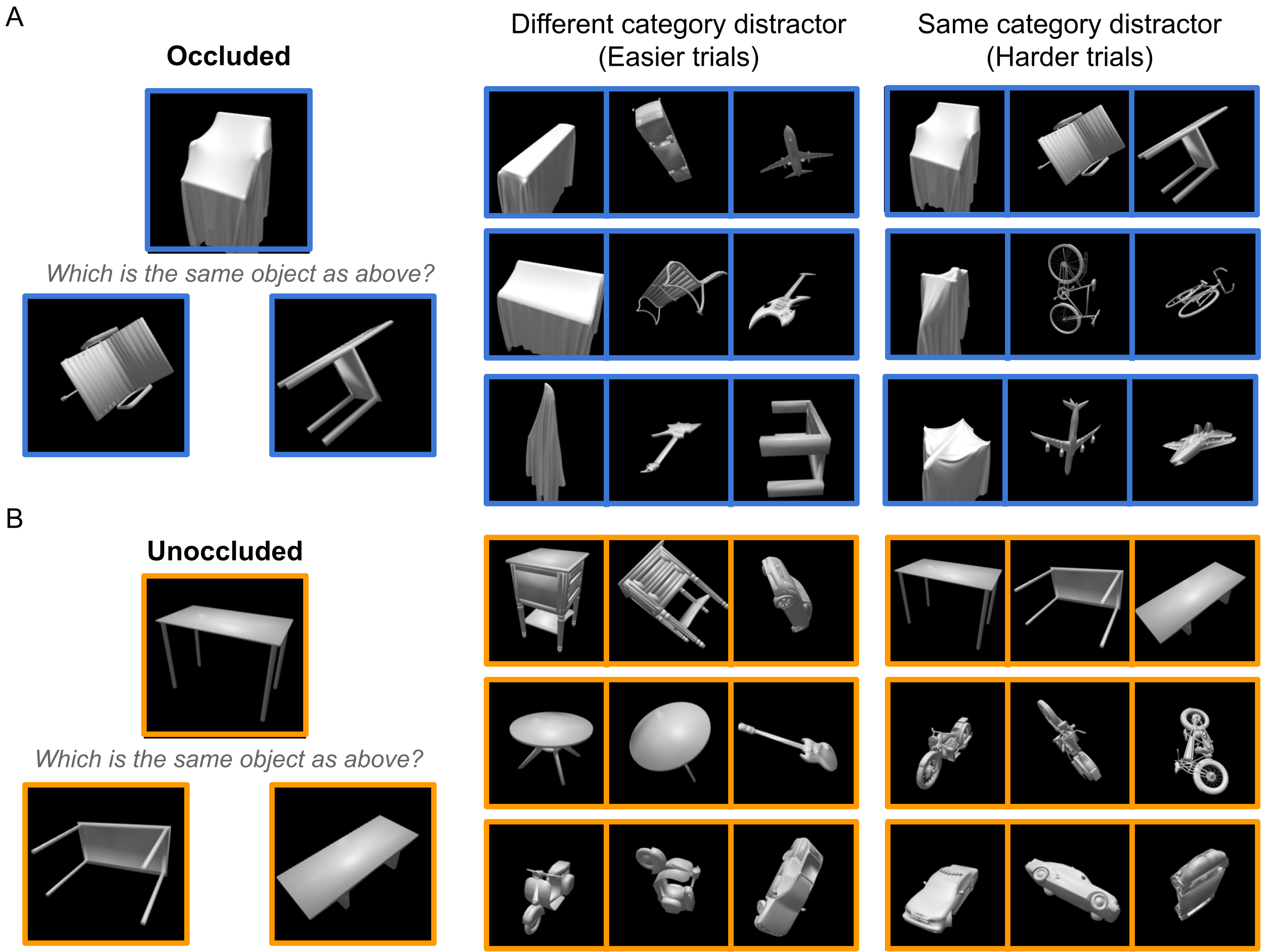}
    \caption{%
    Matching a target shape to one of two unoccluded test objects.
    (A)
    Left: A trial in the occluded task condition. The top image shows a ``target item'', the bottom-left image shows a ``matching test item'' and the bottom-right image shows a ``distractor test item''.
    Right: Trials from the occluded task. Each triplet displays, from left-to-right, target item; matching test item; and distractor test item for one trial. We show ``easier'' trials, with different-category distractor and matching test items, and ``harder'' trials, where both test items are of the same category.
    (B)
    Left: A trial from the unoccluded task condition, spatial configuration as in (A).
    Right: Each triplet shows a trial from the unoccluded task, showing instances of easier and harder trials.}
    \label{fig:task}
\end{figure*}

While in some cases humans can recognize draped objects from a single image (e.g., Fig. \ref{fig:intro-demo}C, D), we chose as our experimental setting a visual match-to-sample task that allows us to directly address the above three possibilities. We choose this setting as we are primarily interested in how generative models can support online, detailed 3D shape perception, rather than object categorization or any kind of memory-based process.  The essence of our proposal is that observers may perceive the 3D shape of cloth-covered objects in arbitrary orientations by approximately simulating in their minds the process of how cloth drapes over the object in three dimensions, and imagining what the resulting 2D image would look like. So we constructed an experimental task that should be directly solvable via this mechanism: 
We show observers an unoccluded matching object along with a target draped shape (i.e., the initial matching object rendered in computer graphics under a simulation of cloth draping) and an unoccluded distractor object, in a two-alternative forced choice (2AFC) task.
We call this the ``occluded'' condition to contrast with a control ``unoccluded'' condition (see below).

We chose 10 different everyday object categories (\textit{airplane, bicycle, bus, car, chair, guitar, motorcycle, pistol, rifle, table}) and sampled object meshes for each category from a large repository of 3D shapes \citep{shapenet2015}. We used 24 unique exemplars from  each category, yielding 120 visual-matching trials; each trial used two shapes, and each unique shape appeared in one trial. Each trial presented an unoccluded target shape, a distractor shape, and the target shape after cloth draping.
We varied the similarity between the distractor and matching items (Fig. \ref{fig:task}A, right) to generate visual-matching triplets ranging in difficulty. In half of the trials, the target and distractor objects were drawn from the same category; these we term ``harder'' trials because same-category shapes are generally more difficult to distinguish than different-category objects, which we call ``easier''.

To create the cloth-occluded stimuli, we simulated cloth draping via a particle-based physics engine \citep{macklin2014unified}; we chose simulation parameters (e.g., number of iterations) and the mechanical and material properties of the simulated cloth (e.g., stiffness and mass) to enable efficient, stable simulation of natural-looking cotton-like cloth (see Materials \& Methods).

In the unoccluded condition, we use the same objects but show the target shape without cloth; see Fig. \ref{fig:task}B. In this version of the task, viewpoint variability and the shape similarity between the matching and distractor test items are the only confounding variables.

\subsection*{Physics-Based Analysis-By-Synthesis (PbAS)}

We formalize the problem of matching a cloth covered object with its unoccluded counterpart as approximate Bayesian inference in a causal generative model. Our physics-based analysis-by-synthesis (PbAS) method combines physics and graphics knowledge with statistical inference and optimization. The model consists of three components: a generative model for scenes and images, feature extraction for approximately Bayesian inference (using a pseudo-likelihood approach), and a simulator-in-the-loop inference engine based on Bayesian optimization. As an account of how people can perceive the shapes of objects under cloth (or other challenging viewing conditions), we posit that each of these three components has some analog in the mind and brain, and that they operate and interact in something like the ways we specify here -- not precisely as we have implemented them, but close enough that the speed and accuracy characteristics of the PbAS model can be quantitatively compared with human behavior, along with different model variants and alternative accounts. 

The generative model in PbAS captures the physical scene variables, including object shape and pose, cloth properties, and the mechanics of how they interact, which together produce the geometry of the occluding cloth surface. 
It further includes a model of graphics -- how surface geometry, material, and light interact to generate an image (some factors, like optics, are handled implicitly; see e.g. \cite{koch2018picture} for an explicit treatment, including modeling, of human representation of visual scene geometry).
Given a hypothesized 3D object shape in a hypothesized pose, the model produces a synthesized or hypothetical image which may be compared with the image actually observed. In the analysis by synthesis framework, perception requires inverting this process to recover the object shape and pose likely to have given rise to the observed image (Fig. \ref{fig:model-figure}A, B).

Like most generative models, PbAS is too complex to invert exactly. A ubiquitous approximation algorithm, Markov Chain Monte Carlo (MCMC), iteratively constructs samples from a target distribution like the posterior, but in our case requires far too many iterations to work because each step includes costly physics simulation.
We sought instead to maximize the posterior using Bayesian optimization \citep{snoek2012practical}, which relative to MCMC provides a guided inference scheme where the next scene hypothesis to evaluate is informed by all (instead of only the current) evaluations of the posterior function \citep{cranmer2020frontier}. BayesOpt simultaneously estimates and optimizes the posterior, providing an algorithm which efficiently samples increasingly more probable hypotheses for object shape and rotation given an input occluded image (Fig.~\ref{fig:model-figure}B).
(Psychologically, BayesOpt can be seen as implementing a kind of goal-conditioned mental imagery; see \cite{hamrick2013mental} for an application in the context of mental rotation.)
The probability of a scene hypothesis is computed by comparing its corresponding rendered hypothesis image with the input, using a feedforward feature hierarchy $f_{enc}$ implemented as the first fully-connected layer of a pretrained DCNN \citep{krizhevsky2012imagenet}. While the goal of inference in our model is posterior probability maximization, the optimization trajectory is also of interest for comparison with human behavior.

PbAS can arrive at a reasonable percept rapidly (Fig.~\ref{fig:model-figure}C, D) compared to sampling-based methods like standard MCMC. It therefore provides a more plausible quantitative standard for understanding average human accuracy, how accuracy improves with longer viewing time, and stimulus-driven variability in response time. 

\subsubsection*{Synthesis: Generative Model}

The generative model consists of (i) latent variables describing the scene: a 3D object shape $S$ and its rotation $R$;
(ii) a forward physics simulator along with cloth parameters: cloth size, position, stiffness, mass, and friction, denoted $f_{\Psi}$; and
(iii) a rendering function and lighting parameters, together denoted $f_{\Gamma}$.
We set the physics simulation parameters $f_{\Psi}$ and renderer parameters $f_{\Gamma}$ to the same values as used for stimuli generation (see Section ``The object-under-cloth task'' and Materials \& Methods).
While the model is designed to perceive cloth-covered objects, it applies to unoccluded objects, as in the unoccluded task condition, as a special case by setting $f_{\Psi}$ to the identity function.

\begin{figure*}[ht!]
    \centering
    \includegraphics[width=1\linewidth]{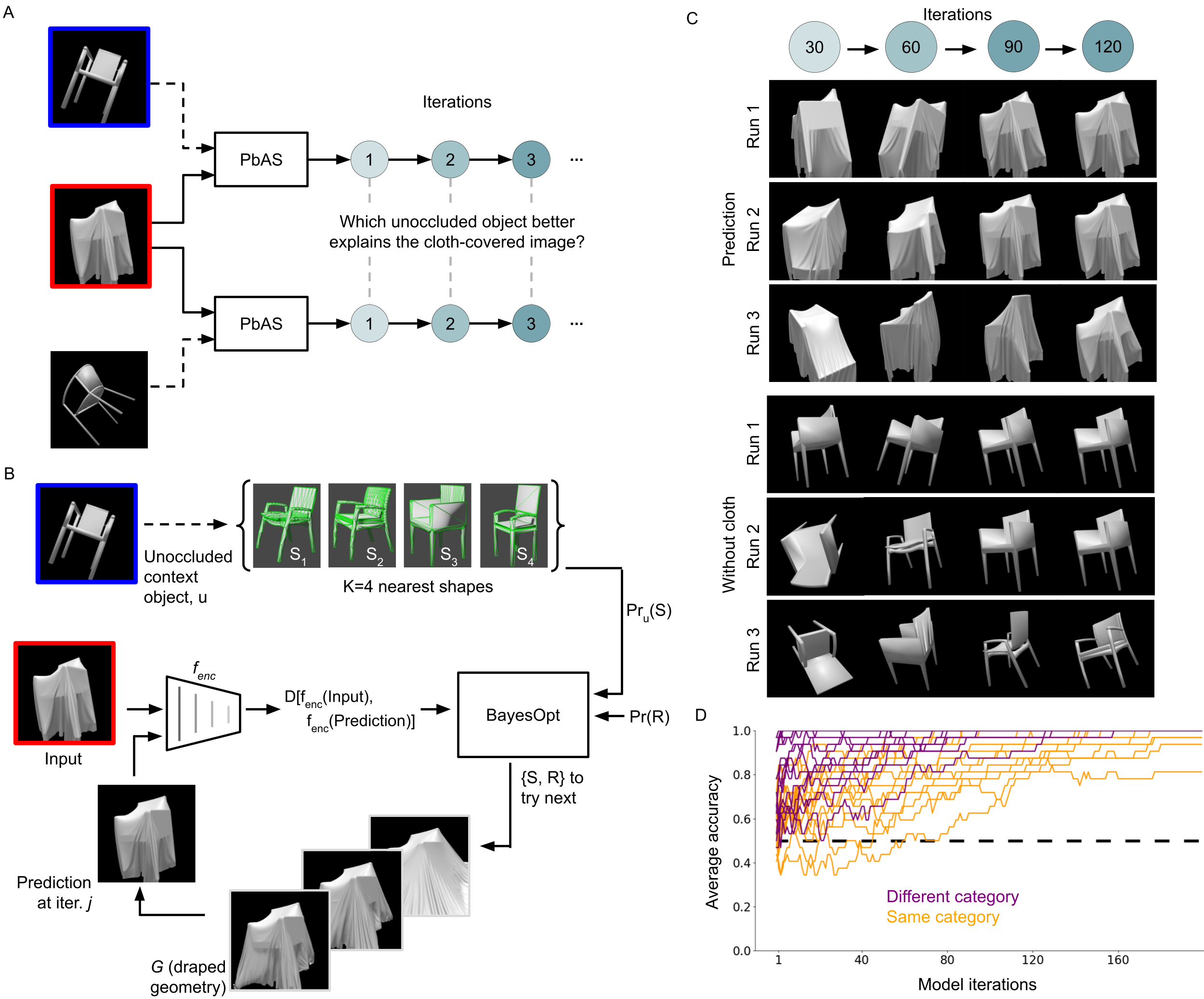}
    \caption{Overview of Physics-based Analysis-by-Synthesis (PbAS).
     (A) Application of the PbAS model to solve the object-under-cloth task. Given an image triplet, two PbAS models are run in parallel; each execution takes as input a test item and the target item. On each iteration, the two executions of the model are compared to determine how well the target item is explained by each test candidate.
     (B) Interpreting an input cloth-covered image (red frame), with a context unoccluded object ($u$; blue frame) supplying a prior over object shape Pr$_u(S)$. Bayesian optimization (BayesOpt) efficiently guides inference, improving shape (S) and rotation (R) hypotheses across iterations. S and R proposals initialize the cloth draping simulation, then are evaluated by computing the distance $D$ between the current scene hypothesis (rendered to a hypothesis image) and the input in a suitable feature encoding space $f_{enc}$.
     (C) Visualization of three inference (panel B) trajectories over time. Rows are independent runs of PbAS each with input as in A, B (red frame) and show the cumulative best scene hypothesis at each iteration. Blocks show hypotheses visualized with (upper; "Prediction") and without (lower; "Without cloth") cloth occlusion. Model estimation accuracy improves with increasing iteration number, but some uncertainty remains as the model (like people) cannot in general perfectly identify the shape or pose of a draped object.
     (D) Evolution of model accuracy averaged across multiple runs in the occluded task condition. Model predictions by iteration for 15 same-category ``harder'' and 15 different-category ``easier'' trials.
    }

    \label{fig:model-figure}
\end{figure*}

Given an occluded input observation (indicated as ``Input'' in red frame, Fig. \ref{fig:model-figure}B) and an unoccluded ``context object'' (in blue frame, Fig. \ref{fig:model-figure}B), we wish to estimate the object shape $S$ and rotation $R$ that best explains the input image. More formally, we wish to invert the generative model to find scene hypotheses that explain perceptual input using Bayesian inference, which amounts to finding the posterior
\begin{align}
  \text{Pr}(S, R \mid I_{obs}) & \propto \text{Pr}(I_{obs} \mid S, R, f_{\Psi}, f_{\Gamma}) \text{Pr}_{u}(S) \text{Pr}(R) \delta_{f_{\Psi}}\delta_{f_{\Gamma}}  \notag \\
&= \text{Pr}(I_{obs} \mid I_{hyp}) \text{Pr}_{u}(S) \text{Pr}(R) \label{bayesrule},
\end{align}
where Pr$( I_{obs}\mid S, R, f_\Psi, f_\Gamma)$ is a likelihood term induced by the physics engine $f_{\Psi}$ and rendering function $f_{\Gamma}$, and the delta functions select fixed physics $f_{\Psi}$ and rendering $f_{\Gamma}$ parameters. For brevity, in the equality in Eq.~\ref{bayesrule} and below we write $I_{hyp}$ $=f_{\Psi}(f_{\Gamma}(S, R))$ for the hypothesis image given latent scene parameters and suppress the delta notation. We next explain each term.

The context object allows observers to form a distribution Pr$_{u}(S)$ over the possible shape of the draped object; because the context object is presented as a 2D rendering, its shape is uncertain.
Even though human observers do not need auxiliary shape information to process cloth-occluded images, this
accompanying context object provides a computationally tractable shape hypothesis space for generative modeling (see the Discussion for future directions relaxing this constraint).
We represent this shape uncertainty Pr$_{u}(S)$ using a categorical distribution over the $K$ nearest neighbors of the actual context object (excluding the context object itself) in a large repository of shapes (the ShapeNet dataset \citep{shapenet2015}; see Fig. \ref{fig:model-figure}B). In our simulations we take $K=4$ and each neighbor is assigned a probability as a function of its distance rank; see Materials \& Methods. We place a uniform prior over rotations Pr$(R)$ covering the half-sphere centered at canonical pose.

Executing the physics simulator $f_{\Psi}$ with a scene hypothesis (a sampled shape $S$ and its rotation $R$) results in a draped cloth geometry $G$ (Fig. \ref{fig:model-figure}B). Passing the resulting scene to the rendering function $f_{\Gamma}$ in turn yields an image $I_{hyp} = f_{\Gamma}(G)$ of the cloth-draped object (Fig. \ref{fig:model-figure}B), i.e. a hypothesis image which may be compared with an input observed image $I_{obs}$ to evaluate its likelihood under the scene hypothesis.

The detailed geometry resulting from cloth simulation (e.g. the particular pattern of wrinkles) can vary significantly with even small changes in the values of random variables \citep{macklin2014unified}; therefore, calculating an accurate likelihood (through marginalization) for any scene hypothesis is computationally intractable \citep{cranmer2020frontier}.
As a result, we define a pseudo-likelihood function Pr$(I_{obs} \mid I_{hyp})$ based on the distance $D(I_{obs}, I_{hyp})$ between the input and hypothesis images in a suitable feature space arising from an encoder $f_{enc}(\cdot)$; here, we set $D = \ell_1$ and adopt the features computed by the first fully-connected layer of AlexNet \citep{krizhevsky2012imagenet} as the encoder $f_{enc}$. The PbAS (pseudo-) likelihood for an input image, given a hypothesis image rendered from a scene proposal, is then
\begin{align*}
    Pr(I_{obs} \mid I_{hyp}) &\propto \exp\left(-\left\| (f_{enc}(I_{obs}) - f_{enc}(I_{hyp}\right) \right\|_{1})
\end{align*}

With these choices for the prior and likelihood, the posterior Pr$(S, R \mid
I_{obs})$ of Eq.~\ref{bayesrule} depends only on two terms: the discrepancy
$Pr(I_{obs} \mid I_{hyp})$ between the observed image and the rendered latent parameters, and the uncertainty $Pr_{u}(S)$ over the shape of the context image.

By measuring the discrepancy between rendered scene hypotheses and observed images in terms of DCNN encoder-based features, the PbAS model as described is an instance of a hybrid top-down/bottom-up (or model-based/cue-based) approach to 3D shape recovery (see also \cite{wang2021neural}). We also consider a purely top-down analysis-by-synthesis approach which is identical except that image discrepancies are computed in terms of raw pixel deviations. The likelihood is then simply
\begin{align*}
 Pr(I_{obs} ms \mid I_{hyp}) \propto \exp\left(-\left\| I_{obs} - I_{hyp} \right\|_{1}\right)
\end{align*}
We refer to this alternative as the ``Pixel Likelihood PbAS'' model, or ``Pixel-PbAS'' for short.

\subsubsection*{Analysis: Inference Using Bayesian Optimization}

The posterior Pr$(S, R \mid I_{obs})$ of Eq.~\ref{bayesrule} contains all information that our model extracts from an observed image $I_{obs}$, but computing this distribution is intractable.
Standard simulation-based inference methods based on MCMC ensure eventual convergence to the full posterior but in practice spend too many iterations in low probability regions \citep{cranmer2020frontier}.
We focus instead on the maximum a posteriori (MAP) setting: finding the best single scene interpretation rather than the full posterior over all possible latent variable settings.
Following previous work in simulation-based inference \citep{jarvenpaa2019efficient,kandasamy2015bayesian,tamura2018bayesian}, we
employ Bayesian optimization (or BayesOpt \citep{snoek2012practical}); unlike gradient-based algorithms, BayesOpt allows us to optimize functions which include procedures, such as our scene renderer, which do not expose or do not support gradients. See Materials \& Methods for an overview and details of BayesOpt applied in PbAS.

\subsubsection*{Solving the Object-Under-Cloth Task Using the Model}

Human participants see two unoccluded context objects (i.e., test items) and one target object on each trial. Recall that by its design, the PbAS model interprets a target (i.e., cloth-draped or unoccluded depending on experiment condition) object in the context of an unoccluded object. Thus, to model a given trial, we form two pairs, each consisting of a context object (either the matching item or the distractor item) and the target object, and apply PbAS to each pair (Fig. \ref{fig:model-figure}A). Each PbAS run aims to explain the same input image, but with different shape hypotheses derived from either the matching object or the distractor object. At every iteration, we save the current best parameter estimates (i.e., shape and rotation) and the log posterior score for that scene hypothesis. Using the odds ratio decision rule, we obtain the model's best estimate of the underlying shape for each inference step.

We ran the PbAS model 32 times on each trial, for 200 iterations each, and treated each of these runs as a simulated participant (although with finer temporal resolution). At each of the 200 iterations, we averaged the binary decisions across runs to obtain mean accuracy predictions -- i.e., simulating the accuracy of participants' average shape choices. In our analysis we compare the dynamics of model choice with human decisions sampled at three different time intervals, corresponding to three different presentation durations that varied across experimental conditions (see the Section ``Iterative refinement in PbAS explains Human Accuracy and Response Time'' for comparisons of models and human behavior). Fig. \ref{fig:model-figure}D shows how the average model performance changes as a function of iteration for a subset of our stimuli.

\subsubsection*{Bottom-Up Models Based On DCNNs}
To help evaluate the PbAS model and its correspondence with human perception, we considered several well-studied bottom-up models as comparisons for human and model performance.
Recent computer vision models based on DCNNs learn powerful visual feature hierarchies achieving state-of-the-art object recognition performance.
These feature hierarchies are relatively robust to variation in pose and lighting, can predict certain aspects of variance in neural and behavioral data, and are considered the ``current best models of the primate visual stream'' \citep{schrimpf2018brain}. Moreover they are useful for visual tasks beyond object recognition; these features have been used for a number of other vision problems e.g. object localization and pose estimation \citep{yamins2016using}, among others, with minor or no modification.
In testing these pretrained models, our goal is not to establish whether DCNNs, considered as a model class, can perform the object-under-cloth task.
DCNNs are universal function approximators; with enough data, enough compute, and the right architecture and optimization procedure, they are likely able to learn to perform our visual-matching task.
Instead, our goal is to assess whether the features learned from categorizing objects in natural scenes can suffice to perceive cloth-occluded shapes as well.

Because our synthesized stimuli and task design differ from those used for the pretrained DCNNs, we also test the same networks after fine-tuning them using images similar to our experimental stimuli.
We tested the following architectures, each pretrained using ImageNet ~\citep{deng2009imagenet}: AlexNet \citep{krizhevsky2012imagenet}, ResNet-50 \citep{he2016deep}, and VGG16 \citep{simonyan2014very}.
Each DCNN was fine-tuned separately for the cloth-occluded and unoccluded conditions. The task was the same visual matching problem presented to humans: given an image containing two unoccluded test shapes and one target object (a ``triplet''; objects sampled from a total of 50 shapes), determine which test shape corresponds to the target.
We repeated this process 32 times; thus we fine-tuned 32 copies (to match the number of PbAS runs per trial) of each architecture for each occlusion condition. We report the average accuracy of these 32 fine-tuned networks. See Materials \& Methods for dataset generation, fine-tuning, and evaluation procedures.

For both the pretrained and fine-tuned conditions, we found that no architecture was more accurate than AlexNet (see Fig. \ref{fig:dcnn-results}).
Therefore, we use both the pretrained AlexNet and our fine-tuned variant in our comparisons of bottom-up models with behavior.
\subsection*{Iterative Refinement in PbAS Explains Human Accuracy and Response Times}

To evaluate PbAS as a candidate model for human perception, we compared its predictions on the object-under-cloth task with two key behavioral measures: average accuracy and response time. 
We recruited human subjects and assigned them to either the occluded or unoccluded condition (see Fig. \ref{fig:task}A, B left panels). Participants were also divided into three presentation time conditions: the two fixed (1 or 2 second) time conditions and the unlimited time condition, which presents stimuli until subjects respond.
In total, the experiment consisted of 2 occlusion $\times$ 3 presentation time = 6 conditions in a between-subjects design.

As is typical in modeling studies, we compared the average accuracy of PbAS and alternative models with that of humans. Because accuracy measures alone might simply favor models that are more performant, we also examined how PbAS ``response times'' -- the number of inference iterations used per trial -- might explain human response times on the same trials.

\subsubsection*{Explaining Human Accuracy Across Presentation Time Conditions}

We first established that behavioral performance is significantly affected by task setting. While participants performed well above chance across all occlusion and presentation time conditions, their performance varied with respect to these design parameters. Most obviously, human performance was better in the unoccluded setting ($p < .05$). With longer presentation time, average performance significantly improved ($p < .05$, Fig. \ref{fig:behavior-results}A; see Fig. \ref{fig:behavior-results}B for results broken down by occlusion condition) and response times increased ($p < .05$, Fig. \ref{fig:behavior-results}C; see Fig. \ref{fig:behavior-results}D for results broken down by occlusion condition). We also note that there was no learning effect throughout the experiment, with participants' average performance remaining fairly constant across trials (Fig. \ref{fig:learning-curves}). 

\begin{figure*}[ht!]
    \centering
    \includegraphics[width=.7\linewidth]{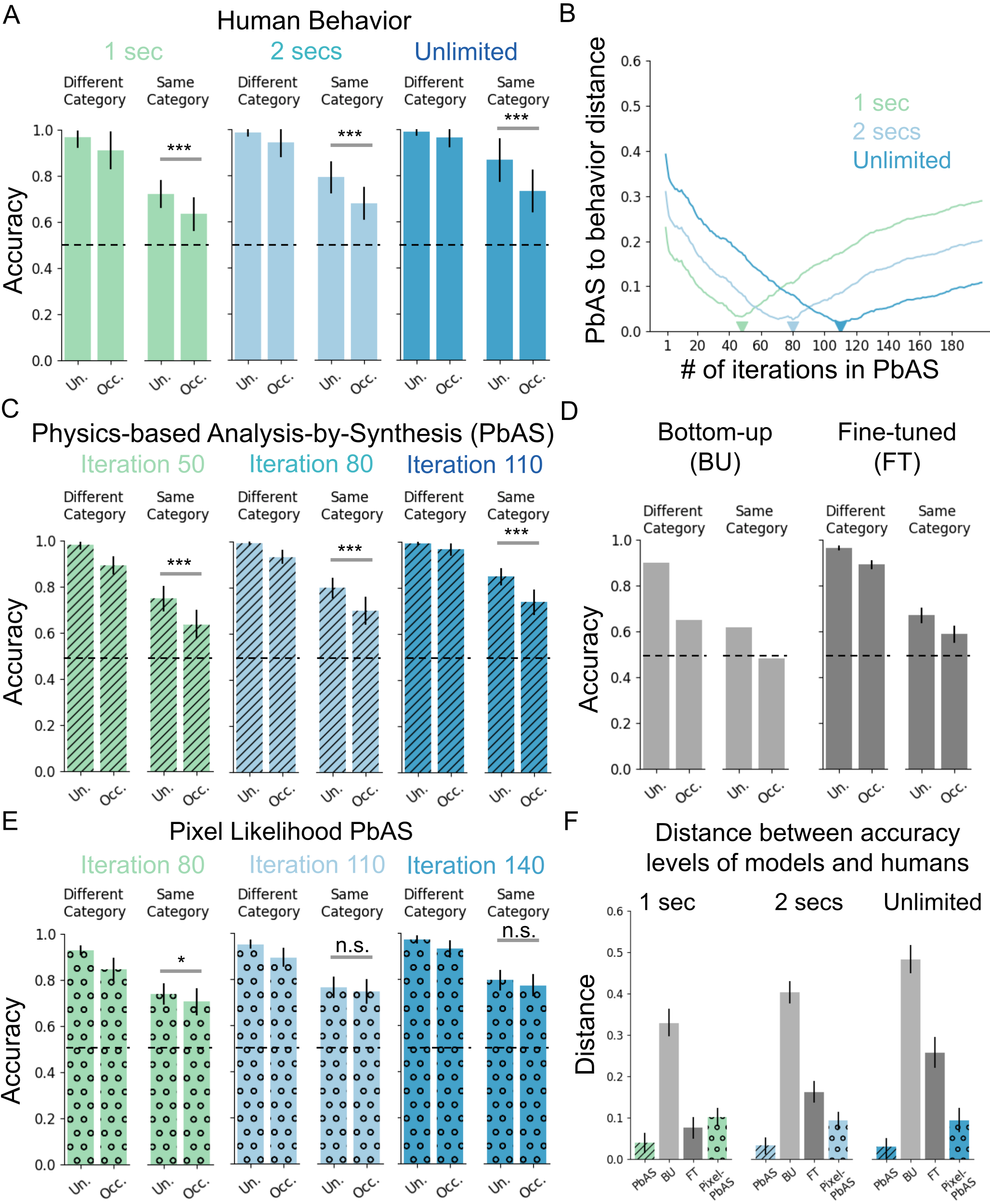}
    \caption{PbAS explains how human accuracy increases with longer stimulus presentation time.
      (A) Behavioral accuracy for each presentation time, occlusion condition, and difficulty. (A trial is said to be hard if the distractor test item is of the same category as the target item, and easy if the distractor test item is of a different category than the target item.)
      (B) Divergence between model and human performance at each model iteration. Colored lines show $\ell_2$ distance between PbAS model and human accuracy levels in indicated presentation time condition. Human accuracy at each increasing  presentation time is best matched by model at correspondingly greater iteration (colored triangles).
      (C) Accuracy of the PbAS model at the three iteration numbers chosen to be close to the best matching iterations marked by colored triangles in (B). (We show results for 50 rather than 48 iterations; see text for details.) Evolution of PbAS accuracy levels over these snapshots closely matches human accuracy levels at the corresponding presentation times; compare (A).
      (D) Performance of the bottom-up network (BU) and the fine-tuned (FT) variants; FT model reports ensemble average. Unlike humans and the PbAS model, in harder cloth-occluded trials with same-category distractors, the BU and FT models remain close to chance (dashed lines). 
      (E) Performance of the ``Pixel-PbAS'' model, which performs a more top-down form of analysis-by-synthesis by attempting to explain input images at the pixel level, by excluding the bottom-up image encoding module. Relative to the PbAS model, this model requires more iterations to reach human-level performance; more critically, it qualitatively misses a key aspect of behavior by performing equally well across occlusion conditions, specifically in the harder same category trials. Error bars in panels A, C and E show standard deviation; significance %
      determined using independent-sample t-tests.
      (F) $\ell_2$ distance between human accuracy (A) and models: PbAS, bottom-up network pretrained (BU) and after fine-tuning (FT), and PbAS without image encoding  (using pixels for likelihood computation; ``Pixel-PbAS''). For the PbAS and Pixel-PbAS models, for each presentation time, we present the distances based on their corresponding best-matching iteration number. 
      Error bars show $95\%$ bootstrapped confidence intervals; ``***'': $p<.001$; ``*'': $p<.05$; ``n.s.'': not significant ($p > .05$). 
      }
    \label{fig:model-behavior-accuracy}
\end{figure*}

The design of our behavioral experiment offers a multifaceted view of human performance in terms of presentation time, trial difficulty (defined as whether test items are of same or different category; Fig. \ref{fig:task}), and occlusion condition.
In Fig. \ref{fig:model-behavior-accuracy}A, we present average human accuracy levels for each presentation time, pooled with respect to the two difficulty types (``Different Category'' vs. ``Same Category'') and two occlusion conditions (occluded vs. unoccluded). 
Observers performed significantly above chance even in the most challenging setting with cloth occlusion, same-category distractors, and the briefest presentation time (1 second).
Note also that performance improved with longer presentation time in the same-category distractor trials where, unlike the easier different-category distractor case, performance does not reach ceiling even with unlimited presentation time. We now ask whether PbAS and bottom-up models can explain these nuanced results. %

\clearpage

We compared average human accuracy levels for each presentation time condition (collapsing over occlusion and difficulty) with PbAS accuracy at each model iteration. The comparison used the $\ell_2$ distance. We found that the longer the presentation time, the more model iterations are needed to best match behavior: the fit for 1 second data requires fewer (48) iterations than are needed for the 2 second condition (80), and even more iterations (110) are needed to match the unlimited time data (Fig. \ref{fig:model-behavior-accuracy}B).
The performance of the PbAS model at the best-fitting iteration numbers for each presentation time closely matches their corresponding behavioral accuracies (compare Fig. \ref{fig:model-behavior-accuracy}A and C, which shows model performance at iterations 50, 80, and 110 for simplicity; model accuracy levels at 48 and 50 iterations are essentially identical). In particular, the correspondence between PbAS and behavior (measured as the $\ell_2$ distance between behavioral and model accuracy levels) is stronger than it is for any other model (Fig. \ref{fig:model-behavior-accuracy}F, $p < .05$, except PbAS vs. FT in the 1 second condition; see below).

Unlike the PbAS model, the bottom-up features derived from pre-trained DCNNs failed to explain human accuracy levels, nor did they after fine-tuning these networks separately for each occlusion condition (Fig. \ref{fig:model-behavior-accuracy}D, F).
As expected, the performance of the pretrained bottom-up network declined substantially under occlusion, but it did so even for the easier different category distractor trials (Fig. \ref{fig:model-behavior-accuracy}D, ``Bottom-up (BU)''). 
For the harder cloth-draped, same-category trials, the performance of the bottom-up model reduced to chance (Fig. \ref{fig:model-behavior-accuracy}D, ``Bottom-up (BU)'').
Fine-tuning this network improved its overall performance, but most of this improvement manifested in the different-category trials and indeed its performance remained near chance in the harder cloth-occluded trials with same category distractors (Fig. \ref{fig:model-behavior-accuracy}D, ``Fine-tuned (FT)'').
These results are reflected in the correspondence between human and network accuracy. In all but one condition, the discrepancy between bottom-up and fine-tuned models, and human behavior, is higher than it is for PbAS (Fig. \ref{fig:model-behavior-accuracy}F). (In the 1-second condition, the fine-tuned model is statistically inseparable from PbAS, but it decouples from behavior in finer-grained trial-by-trial analysis, as we explain in the next section (see also Figs.~\ref{fig:model-behavior-accuracy-finegrained}, ~\ref{fig:supp-correlations}, \ref{fig:supp-correlations-same-category})).

Overall, unlike PbAS, the discrepancy between human and network accuracy levels increased with presentation time, suggesting the need for additional computations beyond the bottom-up processing implemented in these DCNN models
(Fig.~\ref{fig:model-behavior-accuracy}F). 

These results provide support for the role of top-down computations (the generative model) in the hybrid architecture embodied in PbAS: The DCNN feature hierarchies that alone cannot explain behavior are useful when they guide inference (by defining the likelihood) in the generative model. Is this bottom-up component necessary to explain behavior? We evaluated a model that removed the image encoding module. This ablation -- referred to as the Pixel Likelihood PbAS (or ``Pixel-PbAS'' for short) --  computes likelihood in the pixel space, keeping everything else unchanged from PbAS.
We found that this ablation fails to reproduce an important aspect of behavior: Unlike the PbAS model and human judgments, the Pixel-PbAS model performs equally well in the harder (i.e., same category) occluded and unoccluded trials (Fig. \ref{fig:model-behavior-accuracy}E). Moreover, it takes longer to reach human level accuracy relative to PbAS, requiring about 30 more iterations for each presentation time condition (Fig. \ref{fig:model-behavior-accuracy}E). Finally, this model does not match behavior as well as PbAS; using its best-fitting iteration numbers, the distance to behavior is greater than that of PbAS (Fig. \ref{fig:model-behavior-accuracy}F; $p<.05$ for each  pairwise comparison, using direct bootstrap hypothesis testing). However, we note that unlike the bottom-up models, the distance from the Pixel-PbAS model to behavior is constant or decreases slightly across presentation time conditions, indicating that the iterative refinement of scene hypotheses is still crucial to explain how behavioral performance improves with longer exposure times. These results establish that both top-down and bottom-up components of the PbAS architecture are needed to account for behavior. 
Relative to the bottom-up models, PbAS's superior account of behavior is not merely a result of its better task performance, but is instead due to its making similar perceptual judgments, and errors, as humans. The next two sections provide further evidence for these conclusions using fine-grained error and response time analyses. 

\subsubsection*{Explaining Trial-Level Human Accuracy}

\begin{figure*}[ht!]
    \centering
    \includegraphics[width=1\linewidth]{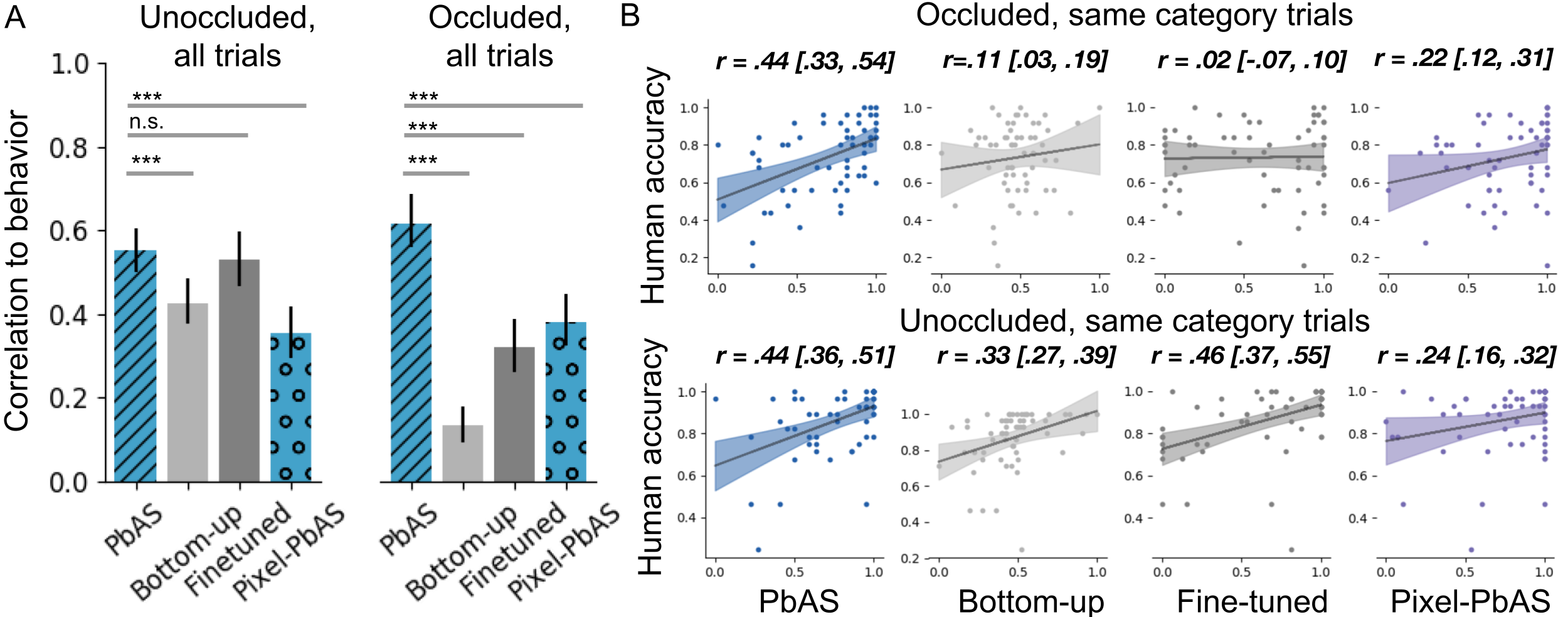}
    \caption{
    Fine-grained analysis of human accuracy at the level of individual trials in the unlimited time condition. PbAS explains behavior better than alternative models.
      (A) Trial-level average accuracy correlations between models and humans in the unlimited time condition. 
      PbAS, bottom-up network pretrained (BU) and after fine-tuning (FT), and the PbAS model without image encoding (``Pixel-PbAS''). (The fine-tuned model reports ensemble average of multiple fine-tuned networks.)
      Error bars show bootstrapped $95\%$ confidence intervals. Statistical comparisons are made using direct bootstrap hypothesis testing (``***'': $p<.001$; ``n.s.'': $p > .05$).
      (B) The hardest, same-category trials reveal that only the PbAS model consistently correlates with behavior in both the unoccluded and occluded conditions. The x-axis values are normalized to range between 0 and 1. Correlation coefficients are indicated on each scatter plot; bootstrapped $95\%$ confidence intervals in brackets.  
      }
    \label{fig:model-behavior-accuracy-finegrained}
\end{figure*}

Next, we evaluated the ability of the models to explain average human accuracy at the level of individual trials.
In the unlimited time condition, we found that the trial-by-trial accuracy of the PbAS model at the best-fitting iteration (iteration 110, marked by the dark blue triangle in Fig. \ref{fig:model-behavior-accuracy}B) correlated well with behavior, and did so consistently in both occlusion conditions ($0.55$ and $0.62$ in unoccluded and cloth-occluded conditions; Fig. \ref{fig:model-behavior-accuracy-finegrained}A).
In the unoccluded condition, PbAS better correlated with behavior relative to the pretrained bottom-up network features ($p<.001$, using bootstrap direct hypothesis testing), but fine-tuning was effective in closing the gap; PbAS and the fine-tuned model showed no difference ($p=.31$).
However, in the occluded condition, the PbAS model better explained behavior relative to both the pretrained and fine-tuned alternatives ($p<.001$; Fig. \ref{fig:model-behavior-accuracy-finegrained}A).
PbAS also correlated with behavior better than the Pixel-PbAS model in both the unoccluded and occluded conditions ($p < .001$; Fig. \ref{fig:model-behavior-accuracy-finegrained}A; see Fig. \ref{fig:supp-correlations} for qualitatively similar results in the other two presentation time conditions.)
Despite the superior quantitative account of PbAS, we note that none of the models considered could explain all of the reproducible variance in the behavioral data.
Split-half correlations across participants (see Materials \& Methods) in the unlimited presentation time condition were around $r=.80$ for both occlusion conditions, significantly higher ($p<.05$) than the correlation achieved by PbAS.

What underlies the PbAS model's ability to consistently account for behavioral accuracy at the trial-level across both occlusion conditions? We hypothesize that both top-down generative knowledge and our bottom-up feature embedding are crucial. 
To address, we first notice that in the easier, different category trials, humans performance is at ceiling,  especially in the unlimited time condition (see the Different Category bars in Fig. \ref{fig:model-behavior-accuracy}A). There is therefore little variance to explain in these easier trials. Thus, we focus on the difficult same-category trials where there is appreciable variance in behavioral accuracy across trials. We find that in these difficult trials, when compared to the bottom-up models, only PbAS can account for behavior in both occlusion conditions (Unoccluded: $r = .44 [.36, .51]$; Occluded: $r = .44 [.33, .54]$, where $[l, u]$ indicates lower/upper $95\%$ confidence intervals). In the regular, unoccluded condition, the fine-tuned model (and to some extent the pretrained model) can explain some of these fine-grained behavioral patterns, however, these models, especially the fine-tuned model, decouple from behavior under cloth occlusion (Fig. \ref{fig:model-behavior-accuracy-finegrained}B). 
The Pixel-PbAS model also falls short of the performance of the full PbAS model in both occlusion conditions (p < .001; Fig. 5B; see Fig. S4 for qualitatively similar results in the other two presentation time conditions), further demonstrating the necessity of both top-down generative knowledge and the bottom-up image embedding for successful prediction of behavior.

\subsubsection*{Explaining Trial-Level Response Times As Iterative Inference}

Our analyses have so far focused on accuracy. 
Here, we analyze human response times to ask whether the time course of inference in PbAS can explain the evolution of observers' perceptual decision-making at the level of individual trials -- how long they decide to view a stimulus before making their choice. Thus, in the unlimited time condition, we compare the number of iterations required for the model to arrive at a decision on a given trial (in a given experimental condition) with the average human response time for that trial. 
To do so, we devised a simple decision rule in the model that applies to individual trials.
At each model iteration, this decision rule
compares the average model accuracy to a criterion set to the average participant accuracy within the trial's condition. 
We record the earliest iteration that PbAS performance exceeds that criterion (or the maximum iteration number, $200$, otherwise) and take it as a predictor for that trial's average response time.
This is akin to a drift-diffusion model\footnote{Unlike standard drift diffusion models, the drift rate and other parameters arise from model inference; no parameters are fit save the criterion.} \citep{usher2001time}  where evidence accumulation naturally arises from the iterative refinement of scene hypotheses in the PbAS model.
The results are response time predictions for each trial of each condition in the experiment.

\begin{figure*}[ht!]
    \centering
    \includegraphics[width=.9\linewidth]{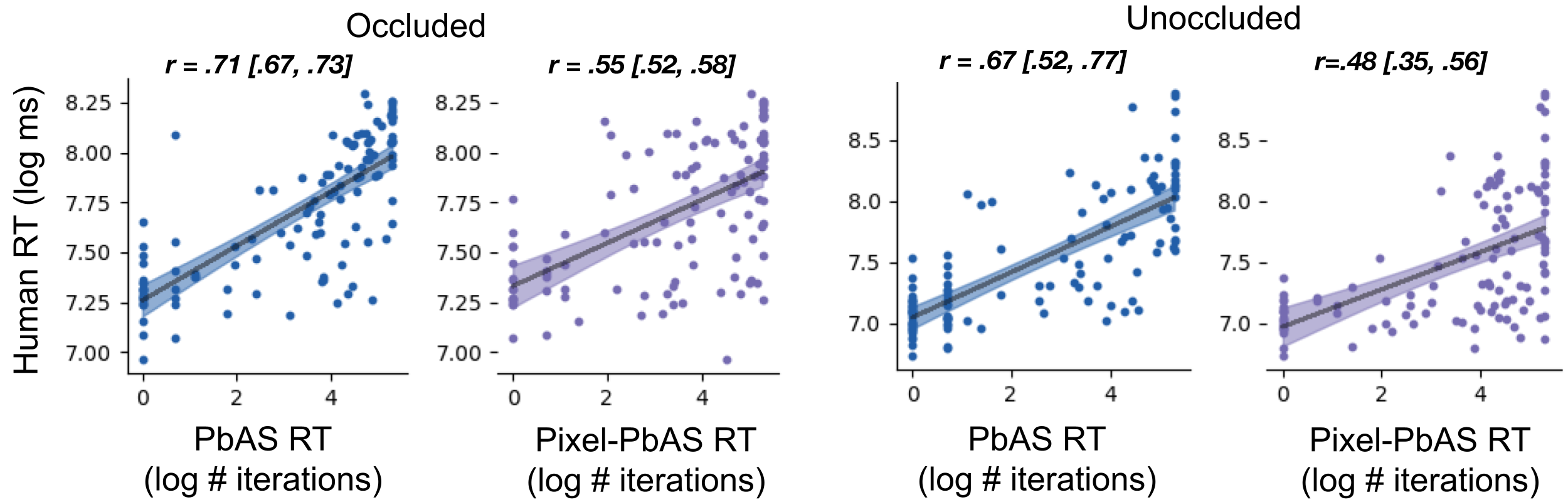}
    \caption{Trial-level response time comparisons.
      Trial-by-trial average human response times (log milliseconds) are explained by the PbAS model (log number of iterations) based on a simple decision threshold (see text for details). The PbAS model captures significantly more variance than the ablated Pixel-PbAS model in each occlusion condition ($p<.001$, using direct bootstrap hypothesis testing). For each comparison, the mean correlation and bootstrapped $95\%$ confidence intervals (in brackets) are shown.
      }
    \label{fig:model-behavior-rt}
\end{figure*} 

Despite the simplicity of this decision rule, we found a remarkable correspondence between the number of iterations needed to solve a trial in PbAS and the time humans took to respond on that trial (Fig. \ref{fig:model-behavior-rt});
the relationship holds for both occlusion conditions. 
No parameters (beyond taking human performance as criterion for each condition) were fit to explain response times.
Because the Pixel-PbAS model also performs iterative inference, we can test its ability to explain response time data as we did with PbAS. %
We found that PbAS gave a better account of response time data than the ablated model in each occlusion condition (Fig. \ref{fig:model-behavior-rt}). %

\section*{Discussion}

We presented evidence for the use of generative model computations in visual perception, in the form of physics-based mental simulations.
Our behavioral results as well as recent related literature \citep{yildirim2016perceiving,phillips2020veiled,little2021physically} raise a fundamental question: How is it possible to perceive the shape of an object when none of the classic visual cues to shape are visible? We proposed that the mind and brain exploit internal representations of the physical processes which form scenes and images.
Our Physics-based Analysis by Synthesis (PbAS) model incorporates knowledge of scene structure and dynamics to explain%
, through online optimization and physics simulation, why a cloth-covered object appears the way it does -- as the result of dropping a cloth on an inferred shape in an inferred pose.
We tested PbAS in a shape matching task which required subjects to match a cloth-draped object with its unoccluded (and randomly rotated) counterpart, in the presence of a distractor.
The PbAS model predicts not only overall human accuracy in this visual matching task, but also how performance improves with longer stimulus presentation times.  Crucially, the number of inference steps needed to reach a behaviorally-determined performance threshold predicts, on a trial-by-trial basis, average participant response times.

Our work adds to the growing literature showing that perception in the brain can be understood as efficient approximate inference in generative models, or analysis-by-synthesis \citep{echeveste2020cortical,yildirim2020efficient,erdogan2017visual,yuille2006vision}.
Past studies have examined some predictions of this theory, but have not provided quantitative evidence that such rich generative models -- %
incorporating shape, object interaction dynamics, and sensory features -- are used online during perception. PbAS also differs from previously considered generative models in its focus on scene elements and causal processes, which when composed allow it to interpret images which are outside typical perceptual experience.
In this way, our work identifies the flexible use of ad hoc dynamic scene properties in perception, such as cloth mechanics, that only indirectly influence image formation and are 
not usually seen as cues to 3D shape. Perceiving shape through cloth occlusion highlights how such ``nuisance'' variables can play a central role in 3D object perception. Our work argues that the compositional use of generative models provides the best way of understanding how these factors influence perception.

Bottom-up models based on DCNNs performed poorly both in the object-under-cloth task and in mimicking human behavior. A DCNN that has been fine-tuned on thousands of images of cloth-occluded objects produces behavior with roughly similar average accuracy as humans in our briefest presentation conditions (1 sec), but unlike the PbAS model fails to explain how performance improves with time and does not correlate at all with trial-by-trial accuracy in the most challenging conditions (occluded with cloth and same-category distractors, for all presentation conditions tested). DCNNs, as a model class, should in principle be able to learn any mapping from inputs to outputs, but our fine-tuning results show that in practice, the data requirements can be substantial (and likely exceed human experience) and the best results far from human-like.
Given the broader context of the many atypical, challenging viewing conditions that the visual system may encounter, these findings underscore the importance of generalization and robustness, ongoing challenges for DCNNs, and illustrate how top-down knowledge can enable perception in difficult novel contexts. Bottom-up models do, however, play an important role in our framework; relative to the ablated Pixel-PbAS model, the hybrid architecture implemented in PbAS demonstrates that powerful feature hierarchies can %
usefully facilitate or guide inference in generative models. This perspective is compatible with much research on ``core'' object recognition showing the explanatory power of bottom-up models \citep{dicarlo2012does,yamins2014performance}.  Future work should also evaluate continuing developments in DCNNs, trained using alternative loss functions, architectures or datasets \citep{konkle2022self, geirhos2021partial}, which may show improved generalization to difficult perceptual tasks.

The PbAS model suggests that perception of cloth-covered objects in the brain relies on a combination of feedforward, feedback, and recurrent computation. We believe that this is valuable as, relative to the case of feedforward processing, there is little evidence to constrain or generate hypotheses regarding the role of feedback and recurrent computation in visual scene analysis \citep{gilbert2013constructive}.
PbAS suggests a new computational goal for feedback and recurrence in the brain, which is in some ways related to pattern theory as expressed in \cite{mumford1994neuronal}:
Such processing might implement the progressive unfolding of one or a number of physical simulations. It is likely that these forms of neural computation implement multiple computational goals needed for such diverse functions as attention, learning, and perception \citep{gilbert2013constructive}. The hypothesis suggested by PbAS  -- internal simulations of physical processes -- is not exclusive of the others and future work should explore their combination.

\begin{figure*}[ht!]
  \label{fig:single_occluded_object}
  \centering
  \includegraphics[width=.75\linewidth]{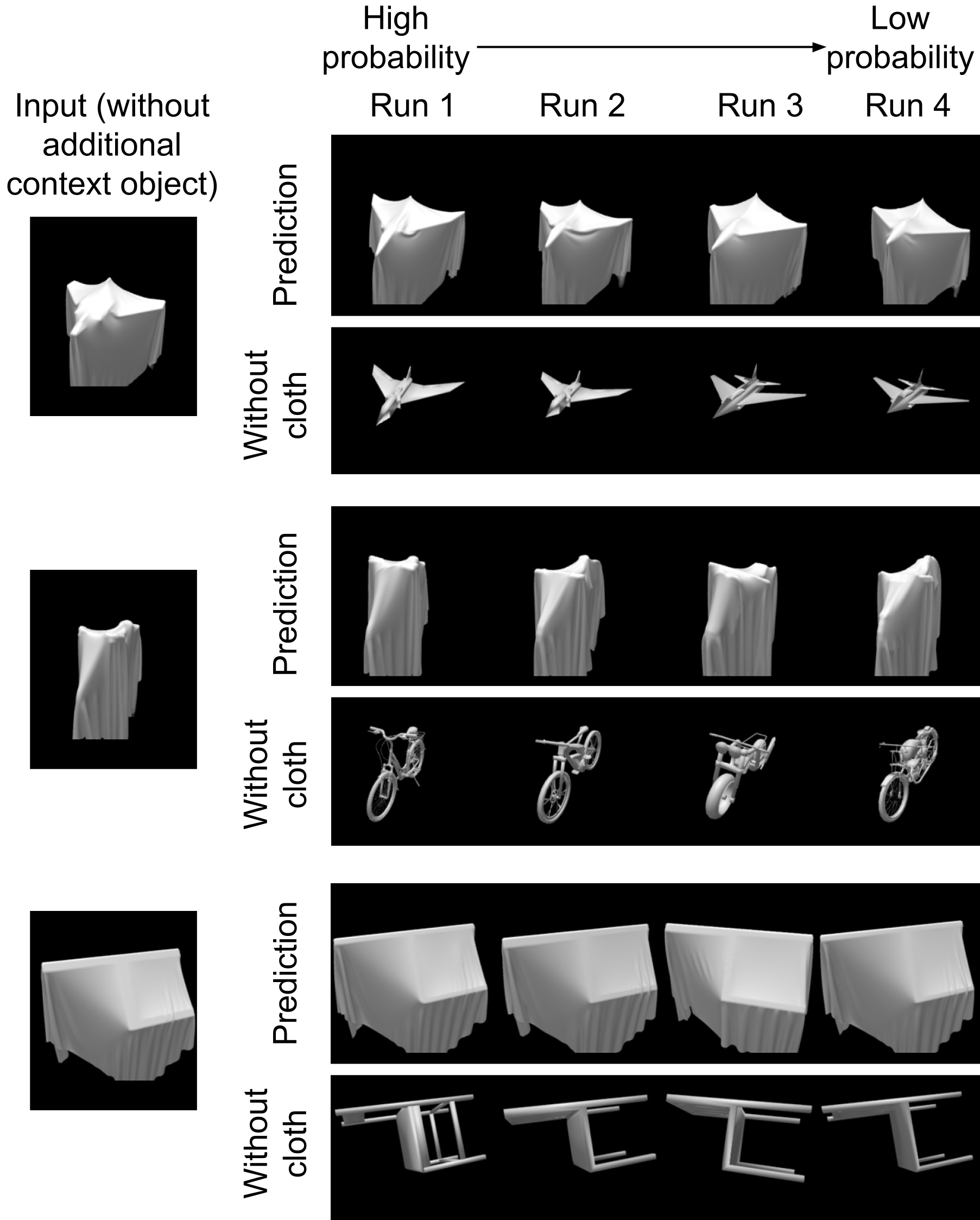}
  \caption{Seeing the shape of a single cloth-draped object, without the aid of unoccluded candidates (cf. Fig.~\ref{fig:task}A, B). By expanding its shape hypothesis space to contain a large set of category-specific objects (as opposed to the four nearest neighbors of the available context object as in our main model) and removing the unoccluded inference module, PbAS can obtain plausible estimates of 3D pose and geometry. Rows show (from left) input image containing target cloth-covered object and four inferred shape/pose hypotheses under this modified PbAS model, ordered from high to low posterior probability.}
  \end{figure*}

The present implementation of the PbAS model accounts for behavior in the specific matching task we studied here (Fig. \ref{fig:intro-demo}A, B; Fig. \ref{fig:task}). Future work should exploit its modular architecture to address other experimental paradigms and perceptual problems.
For example, from an image of a single draped object (without a comparison unoccluded object), humans can often infer its category, approximate pose, and partial shape (Fig. \ref{fig:intro-demo}C-E). While evaluating PbAS in this more difficult scenario is beyond the scope of the present paper, the framework readily extends to this setting; see Fig~\ref{fig:single_occluded_object} for a demonstration from a proof-of-concept implementation.

Our results suggest that shape perception under cloth draping involves mental operations beyond the rapid, bottom-up processing believed characteristic of traditional object recognition \citep{grill2005visual}.
To what extent are the computations hypothesized by PbAS -- 3D shape inference, mental rotation, or mental simulations of physical and image formation processes  -- also engaged in rapid, automatic visual processing? And how do they relate to other cognitive mechanisms supporting dynamic processing such as visual routines \citep{ullman1987visual} and mental imagery \citep{shepard1971mental}?

Recent psychophysical work suggests that these computations %
might be implemented in the visual system as part of spontaneous processing of sensory data. \cite{wong2022seeing} studied cloth-covered object perception across a battery of visual tasks, finding evidence that scenes are rapidly and automatically parsed as the appropriate physical causes. In addition to behavioral probes,
cognitive neuroscience can address where in the brain the computations specified by the PbAS model might be implemented (see \cite{shams2022bayesian} for a recent review of Bayesian causal models and inference in the brain).
For example, fMRI studies \citep{fischer2016functional} have identified brain regions supporting intuitive physical judgments in a dorsal frontoparietal network; it is of significant interest to answer whether the same or similar brain regions are also recruited during the perception of cloth-draped objects.

The PbAS framework discussed and supported here may also play a broader role in visual processing beyond our cloth-draped object setting, unifying competencies beyond traditional shape and object perception. 
A common computational engine may therefore support perception of the dynamical properties of objects, such as the relative masses of colliding rigid bodies or single objects reacting to the application of external forces \citep{sanborn2013reconciling,wu2015galileo,schwettmann2019invariant}; the stiffness of deformable objects undergoing natural transformations \citep{bi2021perception,paulun2017shape,paulun2020visually}; viscosity and flow of liquids \citep{bates2019modeling,kubricht2017consistent,van2018visual};
and in general the perception of the physical (i.e., non-intentional) causal history of an object \citep{chen2016perception,fleming2019getting,schmidt2019visual}. In each of these cases, it is at least plausible that the brain uses generative models to simulate the physical processes that could have produced the observed scene, and compare the results of these simulations to the sensory input. A better understanding of how the brain supports these abilities could also lead to more robust, and more human-like, machine vision systems.

\section*{Materials and Methods}
\subsection*{Generative model}

\subsubsection*{Cloth Simulations}
We used the FLeX engine, a particle-based physics engine, for cloth physics simulation \citep{macklin2014unified}. Simulation parameters as well as the mechanical-material properties of the cloth were chosen so as to achieve fast, stable simulation of natural-looking, cotton-like cloth. Simulation parameters were as follows. Iterations: $4$; subiterations: $19$; particle radius: $0.0078$; collision distance: $0.0078$; shape collision margin:  $0.00078$; particle collision margin: $0.0$; relaxation mode: default; relaxation factor: $1.3$; drag: $0.09$; damping: $0.0$; dissipation: $0.0$; restitution: $0.0$.
The mechanical-material properties of the cloth were as follows. Strength stiffness: $0.8$; bend stiffness: $0.64$; shear stiffness: $0.4$; particle mass: $1.0$; static friction: $0.18$; dynamic friction: $1.1$.

To increase simulation efficiency, we simplified the geometry of the ShapeNet meshes using Blender \citep{blender}).
First, we corrected the surface normals on each mesh by ensuring that they were consistent and pointed outwards. Second, we used
Blender's ``Solidify'' mesh modifier with the thickness parameter set to $-0.0001$. Finally, we merged faces that were adjacent and approximately coplanar (with surface normals differing by less than 0.02 rad $\approx\ang{1.15}$).

We initialized simulations by placing a square cloth (represented computationally with $210 \times 210$ particles)  just above the geometric center of the rotated object to be draped. We then ran the simulation for $150$ steps, sufficient to fully drape all objects we tested.
Each cloth simulation took between $3$ and $40$ seconds on a NVIDIA 2080TI GPU, on the order of $1000$ times faster than alternative implementations using CPU-based cloth simulation and unsimplified meshes.

\subsubsection*{Image Rendering}
The scene was lit to minimize shadows. We placed 14 point lights with energy 0.5 on a sphere with radius 1.22 (object radius normalized to 1), with lights distributed approximately equidistant using the Fibonacci sphere algorithm. We rendered these scenes to $224\times224$ images using Blender's internal renderer.

To equate the texture appearance of the draped and unoccluded images, we replaced the optical materials associated with the original ShapeNet meshes with a diffuse material (diffuse color $0.75$ in each RGB channel, diffuse intensity $0.75$, and specular intensity $0.07$). We used a very similar material to render draped cloths (diffuse color $0.8$ in each RGB channel, diffuse intensity $0.8$, and specular intensity $0.05$). We reasoned that equating the texture appearance in this way would aid the bottom-up neural network models in emphasizing shape over texture \citep{geirhos2018imagenet}.

The experimental stimuli underlying the object-under-cloth task are publicly available: \url{https://github.com/CNCLgithub/intuitive-physics-3d-shape-perception-stimuli}.

\subsubsection*{Approximating Shape Distance}
Given two shapes from ShapeNet $S_i$ and $S_j$, we define a shape distance metric by
(1) rendering each object in a standard canonical pose,
(2) passing each image through a pretrained AlexNet \citep{krizhevsky2012imagenet} and extracting feature activations at the first fully-connected layer (i.e., applying $f_{enc}$ as for pseudo-likelihood evaluation during inference), and
(3) evaluating the $\ell_1$ distance between the feature activations for each shape.
The resulting measure is similar to that used when calculating the pseudo-likelihood.

\subsubsection*{Shape Prior}
Given an unoccluded context object $s_0$, we modeled the observer's shape uncertainty Pr$_u(S)$ as a categorical distribution over the $K=4$ shapes nearest to $s_0$.
Let $d_{s_k}$ be the weight of the $k^{th}$ closest shape $s_k$ to $s_0$; then Pr$_{u}(S = s_k) \propto exp(-d_{s_k})$ with $1 \leq k \leq K$.
The Shapenet database forms a sparse approximation to the space of all object shapes, and we found that the distance between an object and its closest neighbors could vary wildly; one reason is that some object classes have many more exemplars than others. Therefore, a prior defined solely using shape distance showed high variance across trials and was unsuitable for our purposes (e.g., it induces arbitrary bias towards either the distracting or matching object from trial to trial). The unnormalized weights for each nearest shape were instead assigned based on the rank order of their distance to the context object, starting at $d_{s_1} = 750$ and increasing at increments of 75 so that $d_{s_k} = 750 + (k-1) 75$. The scale of these weights was chosen so that the relative contributions of the prior and likelihood were comparable.

\subsection*{Inference Using Bayesian Optimization}

In comparison with traditional inference schemes based on random-walk MCMC \citep{cranmer2020frontier}, Bayesian optimization provides a more guided or ``active'' approach to inference, where the next scene hypothesis to evaluate the posterior on is informed by all of the previous evaluations of the posterior.
In adopting Bayesian optimization, we forego full posterior estimation (which MCMC can provide in principle) in favor of a good point maximum a posteriori (MAP) estimate.
This choice is further motivated by the computational cost of cloth simulation, which is responsible for nearly all of the work our model must do. BayesOpt requires many fewer iterations, and therefore cloth simulations, than random-walk MCMC. It trades expensive overhead (compared to other methods) in choosing search candidates for greater search efficiency \citep{snoek2012practical}.

Following \cite{kandasamy2015bayesian}, we sought to learn a function from latent scene variables (i.e., shape and rotation) to their (unnormalized) log posterior scores. By specifying a tractable Gaussian Process (GP) prior over functions and conditioning on all available data, BayesOpt yields an online strategy for adaptively choosing parameter settings to evaluate and prescribes how the results update the GP posterior. The uncertainty in the GP approximation of the log posterior score decreases as the number of inference iterations increases (i.e., as more evaluations of the posterior are observed).
This probabilistic approximation is computationally cheap to evaluate and has support over the entire range of scene hypotheses (i.e., can be evaluated for any scene hypothesis including those that are previously not evaluated).

BayesOpt requires specification of the GP kernel, which encodes prior assumptions about (e.g) the smoothness of functions \citep{rasmussen2006gaussian}, and an acquisition function which selects the next hypothesis given the results of all previous evaluations. In our work, we used a Matérn kernel with $\nu = 1.5$ (the Matérn $3/2$ kernel) and Automatic Relevance Determination~\citep{rasmussen2006gaussian} to learn a probabilistic mapping from latent scene hypothesis onto posterior scores; and we use the Expected Improvement (EI) as our acquisition function, which favors scene hypotheses that are expected to most improve the posterior score. Each iteration of BayesOpt consists of \textit{(i)} updating the estimated regression function (from scene hypotheses to posterior scores) and \textit{(ii)} optimizing the acquisition function to determine which scene hypothesis to evaluate in the next iteration. We next describe each of these two components.

Scene hypotheses are coded specially for BayesOpt. We represent rotations using normalized Euler angles $R = \{R_x, R_y, R_z\}$, with each orthogonal axis taking values in $[0, 1]$ and together spanning the half-sphere of rotations. The shape variable is discrete, which we transform to a continuous encoding using vector quantization. We map the interval $[0, 0.25)$ to shape hypothesis (i.e., nearest neighbor) 1; $[0.25, 0.50)$ maps to the next nearest neighbor, shape 2, and so on up to shape $4$.
The GP therefore learns a regression function from a 4-dimensional input (three numbers for rotation, one for shape) to a scalar, the log posterior score.

With this GP approximation at hand, we define an ``acquisition function'' which uses the current GP state to select the most promising scene hypothesis to try in the next iteration $j+1$, $(S^{(j+1)}, R^{(j+1)})$.
Various active sampling (or learning) heuristics are proposed in the literature (see e.g. \cite{snoek2012practical,jarvenpaa2019efficient,kandasamy2015bayesian}).
We adopt the EI acquisition function \citep{snoek2012practical}, which chooses the scene hypothesis that is expected to most improve the current posterior score, given all of the previous posterior evaluations.
At each iteration $j$ of our model, the inference procedure evaluates a scene hypothesis chosen to optimize the EI acquisition function. EI uses a parameter, denoted $\epsilon$ (set to $330$ in our simulations), to trade-off between how much to weigh the predicted posterior score vs. the uncertainty around that prediction (notice that the GP-based probabilistic regression provides both the predicted mean posterior score and variance around that prediction for the entire range of scene hypotheses). To find the scene hypothesis that optimizes EI, we generate $100,000$ random scene hypotheses and use the highest scoring to initialize further local search (using L-BFGS-B). This procedure yields the scene hypothesis $(S_j, R_j)$ to be evaluated in the next iteration of the model. We evaluate the posterior at this scene hypothesis using Eq.~\ref{bayesrule}.

We implemented our inference scheme using the BayesOpt \citep{bayesopt} and GPy \citep{gpy2014} packages. Code implementing the PbAS model (as well as our behavioral data and analysis) will be made publicly available before publication.%

\subsection*{Bottom-Up Models}

We tested three DCNN architectures: AlexNet \citep{krizhevsky2012imagenet}, ResNet50 \citep{he2016deep}, and VGG16 \citep{simonyan2014very}. These models provide powerful feature hierarchies that are learned as a result of training to classify images from the large-scale real-world ImageNet \citep{deng2009imagenet} dataset.

\subsubsection*{Imagesets for Fine-tuning}

Imagesets for fine-tuning were derived from the 5 shapes/category $\times$ 10 categories $= 50$ object shapes. These are the identical set of objects as those underlying the experimental training trials used to familiarize human participants with the task.
(We note that in our behavioral experiments, we did not provide feedback during the training phase and indeed did not find any evidence of learning in our behavioral data; see Fig. \ref{fig:learning-curves}.)
We used 8 imagesets per occlusion condition, with 500 unique trials in each set giving $500 \times 8 = 4000$ image triplets. (We evaluated how the amount of data used for fine-tuning influenced performance, finding that performance plateaued at 8 imagesets, compared with alternative groups of 1, 2, 8, 18, 28, and 38 imagesets).
For each triplet, we sampled two objects and randomly rotated, draped, and rendered them using our stimulus generation pipeline. We reserved 2 imagesets for test and the remaining were used for training.
To minimize bias, a set of 8 imagesets was sampled from a larger pool of 54 at the beginning of each fine-tuning procedure.
We fine-tuned each model 32 times for each occlusion condition and report accuracy averaged over the condition-specific replicas.

\subsubsection*{Modifying Network Architectures for Fine-tuning}

To fine-tune AlexNet and VGGG16, we removed their top classification layer and replaced it with a linear fully connected layer of size 120.
We trained the linear layer from scratch and fine-tuned the weights of the layer preceding it.

Unlike AlexNet and VGG16, the ResNet-50 model does not contain multiple final fully-connected layers; thus, we used a modified approach to fine-tune it.%
We replaced both its top classification layer as well as the preceding Average Pooling layer with a convolutional layer with kernel size $2$, stride $2$, and dilation $2$, without zero-padding.
This convolutional layer takes as input $2048$ feature maps (the number of output feature maps in the fourth Residual Block of the ResNet-50 model) each with dimensionality $7\times7$ and outputs $300$ feature maps (each with dimensionality $3\times3$).
The ReLU activation function is applied to the flattened outputs of this convolutional layer, which is followed by a single linear fully-connected layer of size $120$.
We trained the weights of the new convolutional layer as well as the fully-connected layer from scratch, while keeping all other weights in the network unchanged.

\subsubsection*{Details of the Training Procedure}

To adapt the networks to our visual matching task, we used metric learning with a triplet margin loss \citep{schultz2003learning}. The goal is to adapt the network's representational space so that distance in that space reflects the similarity structure of our stimuli. Concretely, the distance between an ``anchor'' image and a ``positive example'' should be smaller than the distance between the anchor and  ``negative example''.
A training triplet has the same structure as our behavioral match-to-sample task setup: anchor corresponds to the target item; positive example corresponds to the ground-truth matching test item; and the negative example corresponds to the distractor test item.
(Remember that the training datasets are crafted differently for each occlusion condition and different networks are trained for each of these occlusion conditions.)
We fine-tuned each architecture for a total of 200 epochs and used a held-out test set to make sure the models did not overfit over the course of training.

We set batch size to $8$ and set the triplet loss margin to $2.0$.
We used the ADAM optimizer \citep{kingma2014adam} with ASMGrad \citep{reddi2018convergence} using the following optimization parameters.
We set $\beta_1$ to $0.9$, $\beta_2$ to $0.999$, learning rate to $1.2\times 10^{-6}$ and $\ell_2$ weight decay to $1.8\times 10^{-3}$ at the beginning of training.
In an attempt to optimize the performance of the bottom-up networks, we explored a range of custom learning rate schedules as well as regularization methods.
During training, we scheduled the learning rate as the following.
The learning rate is multiplied by $1.2$ from epoch 1 to epoch 12.
From epoch 13 to epoch 161, the learning rate is annealed by multiplying it with $0.985$, after which it was kept constant until epoch 200.
In addition, to avoid overfitting, we employ regularization using a weight decay strategy and data augmentation.
From epoch 13 to 161, we multiply the weight decay parameter by $1.04$ and $1.03$ in fine-tuning the occluded and unoccluded task conditions, respectively.
(We observed that without this scheduled weight decay, models essentially memorized the training image set, giving rise to a substantial discrepancy between training and test performance.)
As a form of data augmentation, during training, we randomly perturb each image by adding white noise (variance set to $8.3\times 10^{-3}$) with probability $0.3$ (the added noise was restricted to the foreground pixels). All pixel values were truncated to ensure that their values lie between $0$ and $1$.

\subsubsection*{Evaluation of the bottom-up models on the object-under-cloth task}

The accuracy of the pretrained bottom-up model on a given trial was calculated using the following procedure. Recall that each trial in the object-under-cloth task consists of three images: the target item, the matching test item, and the distractor test item. We compute a feature embedding of each of these three images from the first fully-connected layer of the network. We define a correct answer (accuracy=1 for this trial) from the network if the correlation between the embeddings of the target item and the matching test item (denoted $corr_{m}$) is greater than the correlation between the embeddings of the target item and the distractor test item (denoted $corr_d$). Otherwise, the network got the trial wrong (accuracy=0).
The accuracy levels of the pretrained bottom-up model underlying Fig. \ref{fig:model-behavior-accuracy}D, F are calculated in this way.

In Fig. \ref{fig:model-behavior-accuracy-finegrained} where we require a continuous covariate per trial from each model (as opposed to a binary accuracy label), we use Luce's choice rule (i.e., softmax) to transform the above mentioned correlation values to a continuous score: $corr_m / (corr_m + corr_d)$. Notice that the model predictions in Fig. \ref{fig:model-behavior-accuracy-finegrained}B are normalized to the range of $[0, 1]$ for all models.

The trial-level accuracy of the fine-tuned model is calculated in a manner similar to the PbAS model. For a given trial and a fine-tuned network, we select the test item that is closer to the target item as the network's guess and report the fraction of  correct guesses (i.e., the closest test item was the matching test item) across the ensemble of 32 independently fine-tuned networks.

\subsection*{Behavioral Methods}

\subsubsection*{Participants}

A total of 173 participants were recruited from Amazon's crowdsourcing platform Mechanical Turk. The experiment took about 20 minutes to complete. Each participant was paid $\$1.50$. A total of 12 subjects were excluded due to performing at or below chance performance %
(1 in Unoccluded-1sec; 4 in Occluded-1sec; 3 in Occluded-2secs; and 4 in Occluded-Unlimited).
Approval for our behavioral study was obtained from the Massachusetts Institute of Technology Institutional Review Board (the Committee on the Use of Humans as Experimental Subjects), and we obtained each participant’s informed consent prior to any experimental session.

\subsubsection*{Stimuli and Procedure}

We used 240 unique ShapeNet meshes from 10 object categories to create the 120 match-to-sample shape pairs in our task. We selected 24 objects from each category and allocated them evenly between the same-category (target and distractor from same object category) and different-category conditions, pairing each shape with another from the same category or a different category as appropriate. Pairings were sampled randomly without replacement. We thus obtained 6 same-category and 6 different-category pairs for each object category, with no duplicate shapes across trials.

We designed a visual matching experiment based on the object-under-cloth task. The experiment assigned participants to either the occluded or unoccluded conditions as well as one of three conditions varying presentation time lengths, for a between-subject design with 2 occlusion $\times$ 3 presentation time $=$ 6 conditions.
In the 1- and 2-second conditions, the target and test items were displayed for the indicated period of time and the unlimited time condition let participants view the items for as long as they wished, i.e. until their response. Images appeared and disappeared simultaneously.

The spatial organization of the display differed slightly by occlusion condition.
In the unoccluded condition, the two test images were placed side by side, below the target item; for the occluded condition, the test images were placed side by side but above the target.

Participants completed 10 practice trials before moving on to the 120 experimental trials. Participants were provided with running feedback, seeing their average task performance at every 5th trial throughout the experiment except during the practice block (the performance feedback calculation excluded practice trial accuracy).

\subsubsection*{Split-Half Correlations}
To estimate the data noise ceiling, we used bootstrapped split-half correlations. We sampled $1000$ random splits of our participants in each occlusion condition (only considering the unlimited presentation time condition), each split dividing the participants into two groups of equal size. (Participants were sampled without replacement for each partition.) For a single division (one random split) of participants, we computed the average accuracy of each split-half on each trial, then correlated the group accuracies across all trials. (In essence, we used the responses of one split of participants to model the responses of the other half.) We did the same for each of the $1000$ random splits, yielding $1000$ bootstrap estimates of the behavioral noise ceiling and allowing us to assess their average value and spread. But because this procedure effectively halved our participant number, our split-half correlations likely underestimate the true noise ceiling.

\bibliography{text/references}

\begin{thebibliography}{}
\providecommand{\doi}[1]{\url{https://doi.org/#1}}
\bibcommenthead

\bibitem [\protect \citeauthoryear {%
Bates%
, Yildirim%
, Tenenbaum%
\BCBL {}\ \BBA {} Battaglia%
}{%
Bates%
\ \protect \BOthers {.}}{%
{\protect \APACyear {2019}}%
}]{%
bates2019modeling}
\APACinsertmetastar {%
bates2019modeling}%
\begin{APACrefauthors}%
Bates, C.J.%
, Yildirim, I.%
, Tenenbaum, J.B.%
\BCBL {} Battaglia, P.%
\end{APACrefauthors}%
\unskip\
\newblock
\APACrefYearMonthDay{2019}{}{}.
\newblock
{\BBOQ}\APACrefatitle {Modeling human intuitions about liquid flow with
  particle-based simulation} {Modeling human intuitions about liquid flow with
  particle-based simulation}.{\BBCQ}
\newblock
\APACjournalVolNumPages{PLoS computational biology}{15}{7}{e1007210}.
\newblock

\newblock

\PrintBackRefs{\CurrentBib}

\bibitem [\protect \citeauthoryear {%
Bi%
, Shah%
, Wong%
, Scholl%
\BCBL {}\ \BBA {} Yildirim%
}{%
Bi%
\ \protect \BOthers {.}}{%
{\protect \APACyear {2021}}%
}]{%
bi2021perception}
\APACinsertmetastar {%
bi2021perception}%
\begin{APACrefauthors}%
Bi, W.%
, Shah, A.D.%
, Wong, K.W.%
, Scholl, B.%
\BCBL {} Yildirim, I.%
\end{APACrefauthors}%
\unskip\
\newblock
\APACrefYearMonthDay{2021}{}{}.
\newblock
{\BBOQ}\APACrefatitle {Perception of soft materials relies on physics-based
  object representations: Behavioral and computational evidence} {Perception of
  soft materials relies on physics-based object representations: Behavioral and
  computational evidence}.{\BBCQ}
\newblock
\APACjournalVolNumPages{bioRxiv}{}{}{}.
\newblock

\newblock

\PrintBackRefs{\CurrentBib}

\bibitem [\protect \citeauthoryear {%
{Blender Online Community}%
}{%
{Blender Online Community}%
}{%
{\protect \APACyear {2015}}%
}]{%
blender}
\APACinsertmetastar {%
blender}%
\begin{APACrefauthors}%
{Blender Online Community}%
\end{APACrefauthors}%
\unskip\
\newblock
\APACrefYearMonthDay{2015}{}{}.
\newblock
{\BBOQ}\APACrefatitle {Blender - a 3D modelling and rendering package} {Blender
  - a 3d modelling and rendering package}{\BBCQ}\ [\bibcomputersoftwaremanual].
\newblock
\APACaddressPublisher{Blender Institute, Amsterdam}{}.
\newblock
\begin{APACrefURL} {http://www.blender.org} \end{APACrefURL}
\PrintBackRefs{\CurrentBib}

\bibitem [\protect \citeauthoryear {%
Bulthoff%
}{%
Bulthoff%
}{%
{\protect \APACyear {1991}}%
}]{%
bulthoff1991shape}
\APACinsertmetastar {%
bulthoff1991shape}%
\begin{APACrefauthors}%
Bulthoff, H.%
\end{APACrefauthors}%
\unskip\
\newblock
\APACrefYearMonthDay{1991}{}{}.
\newblock
{\BBOQ}\APACrefatitle {Shape from X: psychophysics and computation} {Shape from
  x: psychophysics and computation}.{\BBCQ}
\newblock
\APACjournalVolNumPages{Computational Models of Visual
  Processing}{}{}{305--330}.
\newblock

\newblock

\PrintBackRefs{\CurrentBib}

\bibitem [\protect \citeauthoryear {%
Chang%
\ \protect \BOthers {.}}{%
Chang%
\ \protect \BOthers {.}}{%
{\protect \APACyear {2015}}%
}]{%
shapenet2015}
\APACinsertmetastar {%
shapenet2015}%
\begin{APACrefauthors}%
Chang, A.X.%
, Funkhouser, T.%
, Guibas, L.%
, Hanrahan, P.%
, Huang, Q.%
, Li, Z.%
\BDBL {}Yu, F.%
\end{APACrefauthors}%
\unskip\
\newblock
\APACrefYearMonthDay{2015}{}{}.
\newblock
\APACrefbtitle {{ShapeNet: An Information-Rich 3D Model Repository}}
  {{ShapeNet: An Information-Rich 3D Model Repository}}\
  \APACbVolEdTR{}{\BTR{}\ \BNUM\ arXiv:1512.03012 [cs.GR]}.
\newblock
\APACaddressInstitution{}{Stanford University --- Princeton University ---
  Toyota Technological Institute at Chicago}.
\PrintBackRefs{\CurrentBib}

\bibitem [\protect \citeauthoryear {%
Chen%
\ \BBA {} Scholl%
}{%
Chen%
\ \BBA {} Scholl%
}{%
{\protect \APACyear {2016}}%
}]{%
chen2016perception}
\APACinsertmetastar {%
chen2016perception}%
\begin{APACrefauthors}%
Chen, Y\BHBI C.%
\BCBT {}\ \BBA {} Scholl, B.J.%
\end{APACrefauthors}%
\unskip\
\newblock
\APACrefYearMonthDay{2016}{}{}.
\newblock
{\BBOQ}\APACrefatitle {The perception of history: Seeing causal history in
  static shapes induces illusory motion perception} {The perception of history:
  Seeing causal history in static shapes induces illusory motion
  perception}.{\BBCQ}
\newblock
\APACjournalVolNumPages{Psychological Science}{27}{6}{923--930}.
\newblock

\newblock

\PrintBackRefs{\CurrentBib}

\bibitem [\protect \citeauthoryear {%
Cranmer%
, Brehmer%
\BCBL {}\ \BBA {} Louppe%
}{%
Cranmer%
\ \protect \BOthers {.}}{%
{\protect \APACyear {2020}}%
}]{%
cranmer2020frontier}
\APACinsertmetastar {%
cranmer2020frontier}%
\begin{APACrefauthors}%
Cranmer, K.%
, Brehmer, J.%
\BCBL {} Louppe, G.%
\end{APACrefauthors}%
\unskip\
\newblock
\APACrefYearMonthDay{2020}{}{}.
\newblock
{\BBOQ}\APACrefatitle {The frontier of simulation-based inference} {The
  frontier of simulation-based inference}.{\BBCQ}
\newblock
\APACjournalVolNumPages{Proceedings of the National Academy of Sciences}{}{}{}.
\newblock

\newblock

\PrintBackRefs{\CurrentBib}

\bibitem [\protect \citeauthoryear {%
Deng%
\ \protect \BOthers {.}}{%
Deng%
\ \protect \BOthers {.}}{%
{\protect \APACyear {2009}}%
}]{%
deng2009imagenet}
\APACinsertmetastar {%
deng2009imagenet}%
\begin{APACrefauthors}%
Deng, J.%
, Dong, W.%
, Socher, R.%
, Li, L\BHBI J.%
, Li, K.%
\BCBL {} Fei-Fei, L.%
\end{APACrefauthors}%
\unskip\
\newblock
\APACrefYearMonthDay{2009}{}{}.
\newblock
{\BBOQ}\APACrefatitle {Imagenet: A large-scale hierarchical image database}
  {Imagenet: A large-scale hierarchical image database}.{\BBCQ}
\newblock
 \APACrefbtitle {Computer Vision and Pattern Recognition, IEEE Conference on}
  {Computer vision and pattern recognition, ieee conference on}\ (\BPGS\
  248--255).
\PrintBackRefs{\CurrentBib}

\bibitem [\protect \citeauthoryear {%
DiCarlo%
, Zoccolan%
\BCBL {}\ \BBA {} Rust%
}{%
DiCarlo%
\ \protect \BOthers {.}}{%
{\protect \APACyear {2012}}%
}]{%
dicarlo2012does}
\APACinsertmetastar {%
dicarlo2012does}%
\begin{APACrefauthors}%
DiCarlo, J.J.%
, Zoccolan, D.%
\BCBL {} Rust, N.C.%
\end{APACrefauthors}%
\unskip\
\newblock
\APACrefYearMonthDay{2012}{}{}.
\newblock
{\BBOQ}\APACrefatitle {How does the brain solve visual object recognition?}
  {How does the brain solve visual object recognition?}{\BBCQ}
\newblock
\APACjournalVolNumPages{Neuron}{73}{3}{415--434}.
\newblock

\newblock

\PrintBackRefs{\CurrentBib}

\bibitem [\protect \citeauthoryear {%
Echeveste%
, Aitchison%
, Hennequin%
\BCBL {}\ \BBA {} Lengyel%
}{%
Echeveste%
\ \protect \BOthers {.}}{%
{\protect \APACyear {2020}}%
}]{%
echeveste2020cortical}
\APACinsertmetastar {%
echeveste2020cortical}%
\begin{APACrefauthors}%
Echeveste, R.%
, Aitchison, L.%
, Hennequin, G.%
\BCBL {} Lengyel, M.%
\end{APACrefauthors}%
\unskip\
\newblock
\APACrefYearMonthDay{2020}{}{}.
\newblock
{\BBOQ}\APACrefatitle {Cortical-like dynamics in recurrent circuits optimized
  for sampling-based probabilistic inference} {Cortical-like dynamics in
  recurrent circuits optimized for sampling-based probabilistic
  inference}.{\BBCQ}
\newblock
\APACjournalVolNumPages{Nature neuroscience}{23}{9}{1138--1149}.
\newblock

\newblock

\PrintBackRefs{\CurrentBib}

\bibitem [\protect \citeauthoryear {%
Erdogan%
\ \BBA {} Jacobs%
}{%
Erdogan%
\ \BBA {} Jacobs%
}{%
{\protect \APACyear {2017}}%
}]{%
erdogan2017visual}
\APACinsertmetastar {%
erdogan2017visual}%
\begin{APACrefauthors}%
Erdogan, G.%
\BCBT {}\ \BBA {} Jacobs, R.A.%
\end{APACrefauthors}%
\unskip\
\newblock
\APACrefYearMonthDay{2017}{}{}.
\newblock
{\BBOQ}\APACrefatitle {Visual Shape Perception as Bayesian Inference of 3D
  Object-Centered Shape Representations.} {Visual shape perception as bayesian
  inference of 3d object-centered shape representations.}{\BBCQ}
\newblock
\APACjournalVolNumPages{Psychological Review}{}{}{}.
\newblock

\newblock

\PrintBackRefs{\CurrentBib}

\bibitem [\protect \citeauthoryear {%
Fischer%
, Mikhael%
, Tenenbaum%
\BCBL {}\ \BBA {} Kanwisher%
}{%
Fischer%
\ \protect \BOthers {.}}{%
{\protect \APACyear {2016}}%
}]{%
fischer2016functional}
\APACinsertmetastar {%
fischer2016functional}%
\begin{APACrefauthors}%
Fischer, J.%
, Mikhael, J.G.%
, Tenenbaum, J.B.%
\BCBL {} Kanwisher, N.%
\end{APACrefauthors}%
\unskip\
\newblock
\APACrefYearMonthDay{2016}{}{}.
\newblock
{\BBOQ}\APACrefatitle {Functional neuroanatomy of intuitive physical inference}
  {Functional neuroanatomy of intuitive physical inference}.{\BBCQ}
\newblock
\APACjournalVolNumPages{Proceedings of the national academy of
  sciences}{113}{34}{E5072--E5081}.
\newblock

\newblock

\PrintBackRefs{\CurrentBib}

\bibitem [\protect \citeauthoryear {%
Fleming%
\ \BBA {} Schmidt%
}{%
Fleming%
\ \BBA {} Schmidt%
}{%
{\protect \APACyear {2019}}%
}]{%
fleming2019getting}
\APACinsertmetastar {%
fleming2019getting}%
\begin{APACrefauthors}%
Fleming, R.W.%
\BCBT {}\ \BBA {} Schmidt, F.%
\end{APACrefauthors}%
\unskip\
\newblock
\APACrefYearMonthDay{2019}{}{}.
\newblock
{\BBOQ}\APACrefatitle {Getting “fumpered”: Classifying objects by what has
  been done to them} {Getting “fumpered”: Classifying objects by what has
  been done to them}.{\BBCQ}
\newblock
\APACjournalVolNumPages{Journal of Vision}{19}{4}{15--15}.
\newblock

\newblock

\PrintBackRefs{\CurrentBib}

\bibitem [\protect \citeauthoryear {%
Geirhos%
\ \protect \BOthers {.}}{%
Geirhos%
\ \protect \BOthers {.}}{%
{\protect \APACyear {2021}}%
}]{%
geirhos2021partial}
\APACinsertmetastar {%
geirhos2021partial}%
\begin{APACrefauthors}%
Geirhos, R.%
, Narayanappa, K.%
, Mitzkus, B.%
, Thieringer, T.%
, Bethge, M.%
, Wichmann, F.A.%
\BCBL {} Brendel, W.%
\end{APACrefauthors}%
\unskip\
\newblock
\APACrefYearMonthDay{2021}{}{}.
\newblock
{\BBOQ}\APACrefatitle {Partial success in closing the gap between human and
  machine vision} {Partial success in closing the gap between human and machine
  vision}.{\BBCQ}
\newblock
\APACjournalVolNumPages{Advances in Neural Information Processing
  Systems}{34}{}{}.
\newblock

\newblock

\PrintBackRefs{\CurrentBib}

\bibitem [\protect \citeauthoryear {%
Geirhos%
\ \protect \BOthers {.}}{%
Geirhos%
\ \protect \BOthers {.}}{%
{\protect \APACyear {2018}}%
}]{%
geirhos2018imagenet}
\APACinsertmetastar {%
geirhos2018imagenet}%
\begin{APACrefauthors}%
Geirhos, R.%
, Rubisch, P.%
, Michaelis, C.%
, Bethge, M.%
, Wichmann, F.A.%
\BCBL {} Brendel, W.%
\end{APACrefauthors}%
\unskip\
\newblock
\APACrefYearMonthDay{2018}{}{}.
\newblock
{\BBOQ}\APACrefatitle {ImageNet-trained CNNs are biased towards texture;
  increasing shape bias improves accuracy and robustness} {Imagenet-trained
  cnns are biased towards texture; increasing shape bias improves accuracy and
  robustness}.{\BBCQ}
\newblock
\APACjournalVolNumPages{arXiv preprint arXiv:1811.12231}{}{}{}.
\newblock

\newblock

\PrintBackRefs{\CurrentBib}

\bibitem [\protect \citeauthoryear {%
Gilbert%
}{%
Gilbert%
}{%
{\protect \APACyear {2013}}%
}]{%
gilbert2013constructive}
\APACinsertmetastar {%
gilbert2013constructive}%
\begin{APACrefauthors}%
Gilbert, C.D.%
\end{APACrefauthors}%
\unskip\
\newblock
\APACrefYearMonthDay{2013}{}{}.
\newblock
{\BBOQ}\APACrefatitle {The constructive nature of visual processing} {The
  constructive nature of visual processing}.{\BBCQ}
\newblock
\APACjournalVolNumPages{Principles of neural science}{5}{}{556--576}.
\newblock

\newblock

\PrintBackRefs{\CurrentBib}

\bibitem [\protect \citeauthoryear {%
{GPy}%
}{%
{GPy}%
}{%
{\protect \APACyear {2012}}%
}]{%
gpy2014}
\APACinsertmetastar {%
gpy2014}%
\begin{APACrefauthors}%
{GPy}%
\end{APACrefauthors}%
\unskip\
\newblock
\APACrefYearMonthDay{2012}{}{}.
\newblock
\APACrefbtitle {{GPy}: A Gaussian process framework in python.} {{GPy}: A
  gaussian process framework in python.}
\newblock
\APAChowpublished {\url{http://github.com/SheffieldML/GPy}}.
\PrintBackRefs{\CurrentBib}

\bibitem [\protect \citeauthoryear {%
Grill-Spector%
\ \BBA {} Kanwisher%
}{%
Grill-Spector%
\ \BBA {} Kanwisher%
}{%
{\protect \APACyear {2005}}%
}]{%
grill2005visual}
\APACinsertmetastar {%
grill2005visual}%
\begin{APACrefauthors}%
Grill-Spector, K.%
\BCBT {}\ \BBA {} Kanwisher, N.%
\end{APACrefauthors}%
\unskip\
\newblock
\APACrefYearMonthDay{2005}{}{}.
\newblock
{\BBOQ}\APACrefatitle {Visual recognition: As soon as you know it is there, you
  know what it is} {Visual recognition: As soon as you know it is there, you
  know what it is}.{\BBCQ}
\newblock
\APACjournalVolNumPages{Psychological Science}{16}{2}{152--160}.
\newblock

\newblock

\PrintBackRefs{\CurrentBib}

\bibitem [\protect \citeauthoryear {%
Hamrick%
\ \BBA {} Griffiths%
}{%
Hamrick%
\ \BBA {} Griffiths%
}{%
{\protect \APACyear {2013}}%
}]{%
hamrick2013mental}
\APACinsertmetastar {%
hamrick2013mental}%
\begin{APACrefauthors}%
Hamrick, J.B.%
\BCBT {}\ \BBA {} Griffiths, T.L.%
\end{APACrefauthors}%
\unskip\
\newblock
\APACrefYearMonthDay{2013}{}{}.
\newblock
{\BBOQ}\APACrefatitle {Mental Rotation as Bayesian Quadrature} {Mental rotation
  as bayesian quadrature}.{\BBCQ}
\newblock
 \APACrefbtitle {NIPS 2013 Workshop on Bayesian Optimization in Theory and
  Practice.} {Nips 2013 workshop on bayesian optimization in theory and
  practice.}
\PrintBackRefs{\CurrentBib}

\bibitem [\protect \citeauthoryear {%
He%
, Zhang%
, Ren%
\BCBL {}\ \BBA {} Sun%
}{%
He%
\ \protect \BOthers {.}}{%
{\protect \APACyear {2016}}%
}]{%
he2016deep}
\APACinsertmetastar {%
he2016deep}%
\begin{APACrefauthors}%
He, K.%
, Zhang, X.%
, Ren, S.%
\BCBL {} Sun, J.%
\end{APACrefauthors}%
\unskip\
\newblock
\APACrefYearMonthDay{2016}{}{}.
\newblock
{\BBOQ}\APACrefatitle {Deep residual learning for image recognition} {Deep
  residual learning for image recognition}.{\BBCQ}
\newblock
 \APACrefbtitle {Proceedings of the IEEE conference on computer vision and
  pattern recognition} {Proceedings of the ieee conference on computer vision
  and pattern recognition}\ (\BPGS\ 770--778).
\PrintBackRefs{\CurrentBib}

\bibitem [\protect \citeauthoryear {%
Hong%
, Yamins%
, Majaj%
\BCBL {}\ \BBA {} DiCarlo%
}{%
Hong%
\ \protect \BOthers {.}}{%
{\protect \APACyear {2016}}%
}]{%
hong2016explicit}
\APACinsertmetastar {%
hong2016explicit}%
\begin{APACrefauthors}%
Hong, H.%
, Yamins, D.L.%
, Majaj, N.J.%
\BCBL {} DiCarlo, J.J.%
\end{APACrefauthors}%
\unskip\
\newblock
\APACrefYearMonthDay{2016}{}{}.
\newblock
{\BBOQ}\APACrefatitle {Explicit information for category-orthogonal object
  properties increases along the ventral stream} {Explicit information for
  category-orthogonal object properties increases along the ventral
  stream}.{\BBCQ}
\newblock
\APACjournalVolNumPages{Nature neuroscience}{19}{4}{613}.
\newblock

\newblock

\PrintBackRefs{\CurrentBib}

\bibitem [\protect \citeauthoryear {%
J{\"a}rvenp{\"a}{\"a}%
\ \protect \BOthers {.}}{%
J{\"a}rvenp{\"a}{\"a}%
\ \protect \BOthers {.}}{%
{\protect \APACyear {2019}}%
}]{%
jarvenpaa2019efficient}
\APACinsertmetastar {%
jarvenpaa2019efficient}%
\begin{APACrefauthors}%
J{\"a}rvenp{\"a}{\"a}, M.%
, Gutmann, M.U.%
, Pleska, A.%
, Vehtari, A.%
, Marttinen, P.%
\BCBL {}\ \BOthersPeriod {.}\end{APACrefauthors}%
\unskip\
\newblock
\APACrefYearMonthDay{2019}{}{}.
\newblock
{\BBOQ}\APACrefatitle {Efficient acquisition rules for model-based approximate
  Bayesian computation} {Efficient acquisition rules for model-based
  approximate bayesian computation}.{\BBCQ}
\newblock
\APACjournalVolNumPages{Bayesian Analysis}{14}{2}{595--622}.
\newblock

\newblock

\PrintBackRefs{\CurrentBib}

\bibitem [\protect \citeauthoryear {%
Kandasamy%
, Schneider%
\BCBL {}\ \BBA {} P{\'o}czos%
}{%
Kandasamy%
\ \protect \BOthers {.}}{%
{\protect \APACyear {2015}}%
}]{%
kandasamy2015bayesian}
\APACinsertmetastar {%
kandasamy2015bayesian}%
\begin{APACrefauthors}%
Kandasamy, K.%
, Schneider, J.%
\BCBL {} P{\'o}czos, B.%
\end{APACrefauthors}%
\unskip\
\newblock
\APACrefYearMonthDay{2015}{}{}.
\newblock
{\BBOQ}\APACrefatitle {Bayesian active learning for posterior estimation}
  {Bayesian active learning for posterior estimation}.{\BBCQ}
\newblock
 \APACrefbtitle {Twenty-Fourth International Joint Conference on Artificial
  Intelligence.} {Twenty-fourth international joint conference on artificial
  intelligence.}
\PrintBackRefs{\CurrentBib}

\bibitem [\protect \citeauthoryear {%
Kingma%
\ \BBA {} Ba%
}{%
Kingma%
\ \BBA {} Ba%
}{%
{\protect \APACyear {2015}}%
}]{%
kingma2014adam}
\APACinsertmetastar {%
kingma2014adam}%
\begin{APACrefauthors}%
Kingma, D.P.%
\BCBT {}\ \BBA {} Ba, J.%
\end{APACrefauthors}%
\unskip\
\newblock
\APACrefYearMonthDay{2015}{}{}.
\newblock
{\BBOQ}\APACrefatitle {Adam: A method for stochastic optimization} {Adam: A
  method for stochastic optimization}.{\BBCQ}
\newblock
\APACjournalVolNumPages{International Conference on Learning
  Representations}{}{}{}.
\newblock

\newblock

\PrintBackRefs{\CurrentBib}

\bibitem [\protect \citeauthoryear {%
Koch%
, Baig%
\BCBL {}\ \BBA {} Zaidi%
}{%
Koch%
\ \protect \BOthers {.}}{%
{\protect \APACyear {2018}}%
}]{%
koch2018picture}
\APACinsertmetastar {%
koch2018picture}%
\begin{APACrefauthors}%
Koch, E.%
, Baig, F.%
\BCBL {} Zaidi, Q.%
\end{APACrefauthors}%
\unskip\
\newblock
\APACrefYearMonthDay{2018}{}{}.
\newblock
{\BBOQ}\APACrefatitle {Picture perception reveals mental geometry of 3D scene
  inferences} {Picture perception reveals mental geometry of 3d scene
  inferences}.{\BBCQ}
\newblock
\APACjournalVolNumPages{Proceedings of the National Academy of
  Sciences}{115}{30}{7807--7812}.
\newblock

\newblock

\PrintBackRefs{\CurrentBib}

\bibitem [\protect \citeauthoryear {%
Konkle%
\ \BBA {} Alvarez%
}{%
Konkle%
\ \BBA {} Alvarez%
}{%
{\protect \APACyear {2022}}%
}]{%
konkle2022self}
\APACinsertmetastar {%
konkle2022self}%
\begin{APACrefauthors}%
Konkle, T.%
\BCBT {}\ \BBA {} Alvarez, G.A.%
\end{APACrefauthors}%
\unskip\
\newblock
\APACrefYearMonthDay{2022}{}{}.
\newblock
{\BBOQ}\APACrefatitle {A self-supervised domain-general learning framework for
  human ventral stream representation} {A self-supervised domain-general
  learning framework for human ventral stream representation}.{\BBCQ}
\newblock
\APACjournalVolNumPages{Nature Communications}{13}{1}{1--12}.
\newblock

\newblock

\PrintBackRefs{\CurrentBib}

\bibitem [\protect \citeauthoryear {%
Krizhevsky%
, Sutskever%
\BCBL {}\ \BBA {} Hinton%
}{%
Krizhevsky%
\ \protect \BOthers {.}}{%
{\protect \APACyear {2012}}%
}]{%
krizhevsky2012imagenet}
\APACinsertmetastar {%
krizhevsky2012imagenet}%
\begin{APACrefauthors}%
Krizhevsky, A.%
, Sutskever, I.%
\BCBL {} Hinton, G.E.%
\end{APACrefauthors}%
\unskip\
\newblock
\APACrefYearMonthDay{2012}{}{}.
\newblock
{\BBOQ}\APACrefatitle {Imagenet classification with deep convolutional neural
  networks} {Imagenet classification with deep convolutional neural
  networks}.{\BBCQ}
\newblock
 \APACrefbtitle {Advances in neural information processing systems} {Advances
  in neural information processing systems}\ (\BPGS\ 1097--1105).
\PrintBackRefs{\CurrentBib}

\bibitem [\protect \citeauthoryear {%
Kubricht%
\ \protect \BOthers {.}}{%
Kubricht%
\ \protect \BOthers {.}}{%
{\protect \APACyear {2017}}%
}]{%
kubricht2017consistent}
\APACinsertmetastar {%
kubricht2017consistent}%
\begin{APACrefauthors}%
Kubricht, J.%
, Zhu, Y.%
, Jiang, C.%
, Terzopoulos, D.%
, Zhu, S\BHBI C.%
\BCBL {} Lu, H.%
\end{APACrefauthors}%
\unskip\
\newblock
\APACrefYearMonthDay{2017}{}{}.
\newblock
{\BBOQ}\APACrefatitle {Consistent Probabilistic Simulation Underlying Human
  Judgment in Substance Dynamics.} {Consistent probabilistic simulation
  underlying human judgment in substance dynamics.}{\BBCQ}
\newblock
 \APACrefbtitle {CogSci.} {Cogsci.}
\PrintBackRefs{\CurrentBib}

\bibitem [\protect \citeauthoryear {%
LeCun%
, Bengio%
\BCBL {}\ \BBA {} Hinton%
}{%
LeCun%
\ \protect \BOthers {.}}{%
{\protect \APACyear {2015}}%
}]{%
lecun2015deep}
\APACinsertmetastar {%
lecun2015deep}%
\begin{APACrefauthors}%
LeCun, Y.%
, Bengio, Y.%
\BCBL {} Hinton, G.%
\end{APACrefauthors}%
\unskip\
\newblock
\APACrefYearMonthDay{2015}{}{}.
\newblock
{\BBOQ}\APACrefatitle {Deep learning} {Deep learning}.{\BBCQ}
\newblock
\APACjournalVolNumPages{Nature}{521}{7553}{436--444}.
\newblock

\newblock

\PrintBackRefs{\CurrentBib}

\bibitem [\protect \citeauthoryear {%
Little%
\ \BBA {} Firestone%
}{%
Little%
\ \BBA {} Firestone%
}{%
{\protect \APACyear {2021}}%
}]{%
little2021physically}
\APACinsertmetastar {%
little2021physically}%
\begin{APACrefauthors}%
Little, P.C.%
\BCBT {}\ \BBA {} Firestone, C.%
\end{APACrefauthors}%
\unskip\
\newblock
\APACrefYearMonthDay{2021}{}{}.
\newblock
{\BBOQ}\APACrefatitle {Physically implied surfaces} {Physically implied
  surfaces}.{\BBCQ}
\newblock
\APACjournalVolNumPages{Psychological Science}{32}{5}{799--808}.
\newblock

\newblock

\PrintBackRefs{\CurrentBib}

\bibitem [\protect \citeauthoryear {%
Liu%
, Knill%
\BCBL {}\ \BBA {} Kersten%
}{%
Liu%
\ \protect \BOthers {.}}{%
{\protect \APACyear {1995}}%
}]{%
liu1995object}
\APACinsertmetastar {%
liu1995object}%
\begin{APACrefauthors}%
Liu, Z.%
, Knill, D.C.%
\BCBL {} Kersten, D.%
\end{APACrefauthors}%
\unskip\
\newblock
\APACrefYearMonthDay{1995}{}{}.
\newblock
{\BBOQ}\APACrefatitle {Object classification for human and ideal observers}
  {Object classification for human and ideal observers}.{\BBCQ}
\newblock
\APACjournalVolNumPages{Vision research}{35}{4}{549--568}.
\newblock

\newblock

\PrintBackRefs{\CurrentBib}

\bibitem [\protect \citeauthoryear {%
Macklin%
, M{\"u}ller%
, Chentanez%
\BCBL {}\ \BBA {} Kim%
}{%
Macklin%
\ \protect \BOthers {.}}{%
{\protect \APACyear {2014}}%
}]{%
macklin2014unified}
\APACinsertmetastar {%
macklin2014unified}%
\begin{APACrefauthors}%
Macklin, M.%
, M{\"u}ller, M.%
, Chentanez, N.%
\BCBL {} Kim, T\BHBI Y.%
\end{APACrefauthors}%
\unskip\
\newblock
\APACrefYearMonthDay{2014}{}{}.
\newblock
{\BBOQ}\APACrefatitle {Unified particle physics for real-time applications}
  {Unified particle physics for real-time applications}.{\BBCQ}
\newblock
\APACjournalVolNumPages{ACM Transactions on Graphics (TOG)}{33}{4}{1--12}.
\newblock

\newblock

\PrintBackRefs{\CurrentBib}

\bibitem [\protect \citeauthoryear {%
Mumford%
}{%
Mumford%
}{%
{\protect \APACyear {1994}}%
}]{%
mumford1994neuronal}
\APACinsertmetastar {%
mumford1994neuronal}%
\begin{APACrefauthors}%
Mumford, D.%
\end{APACrefauthors}%
\unskip\
\newblock
\APACrefYearMonthDay{1994}{}{}.
\newblock
{\BBOQ}\APACrefatitle {Neuronal architectures for pattern-theoretic problems}
  {Neuronal architectures for pattern-theoretic problems}.{\BBCQ}
\newblock
\APACjournalVolNumPages{Large-scale neuronal theories of the
  brain}{}{}{125--152}.
\newblock

\newblock

\PrintBackRefs{\CurrentBib}

\bibitem [\protect \citeauthoryear {%
Nogueira%
}{%
Nogueira%
}{%
{\protect \APACyear {2014}}%
}]{%
bayesopt}
\APACinsertmetastar {%
bayesopt}%
\begin{APACrefauthors}%
Nogueira, F.%
\end{APACrefauthors}%
\unskip\
\newblock
\APACrefYearMonthDay{2014}{}{}.
\newblock
\APACrefbtitle {{Bayesian Optimization}: Open source constrained global
  optimization tool for {Python}.} {{Bayesian Optimization}: Open source
  constrained global optimization tool for {Python}.}
\newblock
\begin{APACrefURL} {https://github.com/fmfn/BayesianOptimization}
  \end{APACrefURL}
\PrintBackRefs{\CurrentBib}

\bibitem [\protect \citeauthoryear {%
Paulun%
\ \BBA {} Fleming%
}{%
Paulun%
\ \BBA {} Fleming%
}{%
{\protect \APACyear {2020}}%
}]{%
paulun2020visually}
\APACinsertmetastar {%
paulun2020visually}%
\begin{APACrefauthors}%
Paulun, V.C.%
\BCBT {}\ \BBA {} Fleming, R.W.%
\end{APACrefauthors}%
\unskip\
\newblock
\APACrefYearMonthDay{2020}{}{}.
\newblock
{\BBOQ}\APACrefatitle {Visually inferring elasticity from the motion trajectory
  of bouncing cubes} {Visually inferring elasticity from the motion trajectory
  of bouncing cubes}.{\BBCQ}
\newblock
\APACjournalVolNumPages{Journal of Vision}{20}{6}{6--6}.
\newblock

\newblock

\PrintBackRefs{\CurrentBib}

\bibitem [\protect \citeauthoryear {%
Paulun%
, Schmidt%
, van Assen%
\BCBL {}\ \BBA {} Fleming%
}{%
Paulun%
\ \protect \BOthers {.}}{%
{\protect \APACyear {2017}}%
}]{%
paulun2017shape}
\APACinsertmetastar {%
paulun2017shape}%
\begin{APACrefauthors}%
Paulun, V.C.%
, Schmidt, F.%
, van Assen, J.J.R.%
\BCBL {} Fleming, R.W.%
\end{APACrefauthors}%
\unskip\
\newblock
\APACrefYearMonthDay{2017}{}{}.
\newblock
{\BBOQ}\APACrefatitle {Shape, motion, and optical cues to stiffness of elastic
  objects} {Shape, motion, and optical cues to stiffness of elastic
  objects}.{\BBCQ}
\newblock
\APACjournalVolNumPages{Journal of vision}{17}{1}{20--20}.
\newblock

\newblock

\PrintBackRefs{\CurrentBib}

\bibitem [\protect \citeauthoryear {%
Phillips%
\ \BBA {} Fleming%
}{%
Phillips%
\ \BBA {} Fleming%
}{%
{\protect \APACyear {2020}}%
}]{%
phillips2020veiled}
\APACinsertmetastar {%
phillips2020veiled}%
\begin{APACrefauthors}%
Phillips, F.%
\BCBT {}\ \BBA {} Fleming, R.W.%
\end{APACrefauthors}%
\unskip\
\newblock
\APACrefYearMonthDay{2020}{}{}.
\newblock
{\BBOQ}\APACrefatitle {The Veiled Virgin illustrates visual segmentation of
  shape by cause} {The veiled virgin illustrates visual segmentation of shape
  by cause}.{\BBCQ}
\newblock
\APACjournalVolNumPages{Proceedings of the National Academy of
  Sciences}{117}{21}{11735--11743}.
\newblock

\newblock

\PrintBackRefs{\CurrentBib}

\bibitem [\protect \citeauthoryear {%
Rasmussen%
\ \BBA {} Williams%
}{%
Rasmussen%
\ \BBA {} Williams%
}{%
{\protect \APACyear {2006}}%
}]{%
rasmussen2006gaussian}
\APACinsertmetastar {%
rasmussen2006gaussian}%
\begin{APACrefauthors}%
Rasmussen, C.E.%
\BCBT {}\ \BBA {} Williams, C.K.%
\end{APACrefauthors}%
\unskip\
\newblock
\APACrefYear{2006}.
\newblock
\APACrefbtitle {Gaussian processes for machine learning} {Gaussian processes
  for machine learning}.
\newblock
\APACaddressPublisher{}{MIT press Cambridge, MA}.
\PrintBackRefs{\CurrentBib}

\bibitem [\protect \citeauthoryear {%
Reddi%
, Kale%
\BCBL {}\ \BBA {} Kumar%
}{%
Reddi%
\ \protect \BOthers {.}}{%
{\protect \APACyear {2018}}%
}]{%
reddi2018convergence}
\APACinsertmetastar {%
reddi2018convergence}%
\begin{APACrefauthors}%
Reddi, S.J.%
, Kale, S.%
\BCBL {} Kumar, S.%
\end{APACrefauthors}%
\unskip\
\newblock
\APACrefYearMonthDay{2018}{}{}.
\newblock
{\BBOQ}\APACrefatitle {On the Convergence of Adam and Beyond} {On the
  convergence of adam and beyond}.{\BBCQ}
\newblock
 \APACrefbtitle {International Conference on Learning Representations.}
  {International conference on learning representations.}
\PrintBackRefs{\CurrentBib}

\bibitem [\protect \citeauthoryear {%
Sanborn%
, Mansinghka%
\BCBL {}\ \BBA {} Griffiths%
}{%
Sanborn%
\ \protect \BOthers {.}}{%
{\protect \APACyear {2013}}%
}]{%
sanborn2013reconciling}
\APACinsertmetastar {%
sanborn2013reconciling}%
\begin{APACrefauthors}%
Sanborn, A.N.%
, Mansinghka, V.K.%
\BCBL {} Griffiths, T.L.%
\end{APACrefauthors}%
\unskip\
\newblock
\APACrefYearMonthDay{2013}{}{}.
\newblock
{\BBOQ}\APACrefatitle {Reconciling intuitive physics and Newtonian mechanics
  for colliding objects.} {Reconciling intuitive physics and newtonian
  mechanics for colliding objects.}{\BBCQ}
\newblock
\APACjournalVolNumPages{Psychological review}{120}{2}{411}.
\newblock

\newblock

\PrintBackRefs{\CurrentBib}

\bibitem [\protect \citeauthoryear {%
Schmidt%
, Phillips%
\BCBL {}\ \BBA {} Fleming%
}{%
Schmidt%
\ \protect \BOthers {.}}{%
{\protect \APACyear {2019}}%
}]{%
schmidt2019visual}
\APACinsertmetastar {%
schmidt2019visual}%
\begin{APACrefauthors}%
Schmidt, F.%
, Phillips, F.%
\BCBL {} Fleming, R.W.%
\end{APACrefauthors}%
\unskip\
\newblock
\APACrefYearMonthDay{2019}{}{}.
\newblock
{\BBOQ}\APACrefatitle {Visual perception of shape-transforming processes:
  'Shape scission'} {Visual perception of shape-transforming processes: 'shape
  scission'}.{\BBCQ}
\newblock
\APACjournalVolNumPages{Cognition}{189}{}{167--180}.
\newblock

\newblock

\PrintBackRefs{\CurrentBib}

\bibitem [\protect \citeauthoryear {%
Schrimpf%
\ \protect \BOthers {.}}{%
Schrimpf%
\ \protect \BOthers {.}}{%
{\protect \APACyear {2018}}%
}]{%
schrimpf2018brain}
\APACinsertmetastar {%
schrimpf2018brain}%
\begin{APACrefauthors}%
Schrimpf, M.%
, Kubilius, J.%
, Hong, H.%
, Majaj, N.J.%
, Rajalingham, R.%
, Issa, E.B.%
\BDBL {}others%
\end{APACrefauthors}%
\unskip\
\newblock
\APACrefYearMonthDay{2018}{}{}.
\newblock
{\BBOQ}\APACrefatitle {Brain-score: Which artificial neural network for object
  recognition is most brain-like?} {Brain-score: Which artificial neural
  network for object recognition is most brain-like?}{\BBCQ}
\newblock
\APACjournalVolNumPages{BioRxiv}{}{}{407007}.
\newblock

\newblock

\PrintBackRefs{\CurrentBib}

\bibitem [\protect \citeauthoryear {%
Schultz%
\ \BBA {} Joachims%
}{%
Schultz%
\ \BBA {} Joachims%
}{%
{\protect \APACyear {2003}}%
}]{%
schultz2003learning}
\APACinsertmetastar {%
schultz2003learning}%
\begin{APACrefauthors}%
Schultz, M.%
\BCBT {}\ \BBA {} Joachims, T.%
\end{APACrefauthors}%
\unskip\
\newblock
\APACrefYearMonthDay{2003}{}{}.
\newblock
{\BBOQ}\APACrefatitle {Learning a distance metric from relative comparisons}
  {Learning a distance metric from relative comparisons}.{\BBCQ}
\newblock
\APACjournalVolNumPages{Advances in neural information processing
  systems}{16}{}{41--48}.
\newblock

\newblock

\PrintBackRefs{\CurrentBib}

\bibitem [\protect \citeauthoryear {%
Schwettmann%
, Tenenbaum%
\BCBL {}\ \BBA {} Kanwisher%
}{%
Schwettmann%
\ \protect \BOthers {.}}{%
{\protect \APACyear {2019}}%
}]{%
schwettmann2019invariant}
\APACinsertmetastar {%
schwettmann2019invariant}%
\begin{APACrefauthors}%
Schwettmann, S.%
, Tenenbaum, J.B.%
\BCBL {} Kanwisher, N.%
\end{APACrefauthors}%
\unskip\
\newblock
\APACrefYearMonthDay{2019}{}{}.
\newblock
{\BBOQ}\APACrefatitle {Invariant representations of mass in the human brain}
  {Invariant representations of mass in the human brain}.{\BBCQ}
\newblock
\APACjournalVolNumPages{Elife}{8}{}{e46619}.
\newblock

\newblock

\PrintBackRefs{\CurrentBib}

\bibitem [\protect \citeauthoryear {%
Shams%
\ \BBA {} Beierholm%
}{%
Shams%
\ \BBA {} Beierholm%
}{%
{\protect \APACyear {2022}}%
}]{%
shams2022bayesian}
\APACinsertmetastar {%
shams2022bayesian}%
\begin{APACrefauthors}%
Shams, L.%
\BCBT {}\ \BBA {} Beierholm, U.%
\end{APACrefauthors}%
\unskip\
\newblock
\APACrefYearMonthDay{2022}{}{}.
\newblock
{\BBOQ}\APACrefatitle {Bayesian causal inference: a unifying neuroscience
  theory} {Bayesian causal inference: a unifying neuroscience theory}.{\BBCQ}
\newblock
\APACjournalVolNumPages{Neuroscience \& Biobehavioral Reviews}{}{}{104619}.
\newblock

\newblock

\PrintBackRefs{\CurrentBib}

\bibitem [\protect \citeauthoryear {%
Shepard%
\ \BBA {} Metzler%
}{%
Shepard%
\ \BBA {} Metzler%
}{%
{\protect \APACyear {1971}}%
}]{%
shepard1971mental}
\APACinsertmetastar {%
shepard1971mental}%
\begin{APACrefauthors}%
Shepard, R.N.%
\BCBT {}\ \BBA {} Metzler, J.%
\end{APACrefauthors}%
\unskip\
\newblock
\APACrefYearMonthDay{1971}{}{}.
\newblock
{\BBOQ}\APACrefatitle {Mental rotation of three-dimensional objects} {Mental
  rotation of three-dimensional objects}.{\BBCQ}
\newblock
\APACjournalVolNumPages{Science}{171}{3972}{701--703}.
\newblock

\newblock

\PrintBackRefs{\CurrentBib}

\bibitem [\protect \citeauthoryear {%
Simonyan%
\ \BBA {} Zisserman%
}{%
Simonyan%
\ \BBA {} Zisserman%
}{%
{\protect \APACyear {2014}}%
}]{%
simonyan2014very}
\APACinsertmetastar {%
simonyan2014very}%
\begin{APACrefauthors}%
Simonyan, K.%
\BCBT {}\ \BBA {} Zisserman, A.%
\end{APACrefauthors}%
\unskip\
\newblock
\APACrefYearMonthDay{2014}{}{}.
\newblock
{\BBOQ}\APACrefatitle {Very deep convolutional networks for large-scale image
  recognition} {Very deep convolutional networks for large-scale image
  recognition}.{\BBCQ}
\newblock
\APACjournalVolNumPages{arXiv preprint arXiv:1409.1556}{}{}{}.
\newblock

\newblock

\PrintBackRefs{\CurrentBib}

\bibitem [\protect \citeauthoryear {%
Snoek%
, Larochelle%
\BCBL {}\ \BBA {} Adams%
}{%
Snoek%
\ \protect \BOthers {.}}{%
{\protect \APACyear {2012}}%
}]{%
snoek2012practical}
\APACinsertmetastar {%
snoek2012practical}%
\begin{APACrefauthors}%
Snoek, J.%
, Larochelle, H.%
\BCBL {} Adams, R.P.%
\end{APACrefauthors}%
\unskip\
\newblock
\APACrefYearMonthDay{2012}{}{}.
\newblock
{\BBOQ}\APACrefatitle {Practical bayesian optimization of machine learning
  algorithms} {Practical bayesian optimization of machine learning
  algorithms}.{\BBCQ}
\newblock
 \APACrefbtitle {Advances in neural information processing systems} {Advances
  in neural information processing systems}\ (\BPGS\ 2951--2959).
\PrintBackRefs{\CurrentBib}

\bibitem [\protect \citeauthoryear {%
Tamura%
\ \BBA {} Hukushima%
}{%
Tamura%
\ \BBA {} Hukushima%
}{%
{\protect \APACyear {2018}}%
}]{%
tamura2018bayesian}
\APACinsertmetastar {%
tamura2018bayesian}%
\begin{APACrefauthors}%
Tamura, R.%
\BCBT {}\ \BBA {} Hukushima, K.%
\end{APACrefauthors}%
\unskip\
\newblock
\APACrefYearMonthDay{2018}{}{}.
\newblock
{\BBOQ}\APACrefatitle {Bayesian optimization for computationally extensive
  probability distributions} {Bayesian optimization for computationally
  extensive probability distributions}.{\BBCQ}
\newblock
\APACjournalVolNumPages{PloS one}{13}{3}{e0193785}.
\newblock

\newblock

\PrintBackRefs{\CurrentBib}

\bibitem [\protect \citeauthoryear {%
Ullman%
}{%
Ullman%
}{%
{\protect \APACyear {1987}}%
}]{%
ullman1987visual}
\APACinsertmetastar {%
ullman1987visual}%
\begin{APACrefauthors}%
Ullman, S.%
\end{APACrefauthors}%
\unskip\
\newblock
\APACrefYearMonthDay{1987}{}{}.
\newblock
{\BBOQ}\APACrefatitle {Visual routines} {Visual routines}.{\BBCQ}
\newblock
 \APACrefbtitle {Readings in computer vision} {Readings in computer vision}\
  (\BPGS\ 298--328).
\newblock
\APACaddressPublisher{}{Elsevier}.
\PrintBackRefs{\CurrentBib}

\bibitem [\protect \citeauthoryear {%
Usher%
\ \BBA {} McClelland%
}{%
Usher%
\ \BBA {} McClelland%
}{%
{\protect \APACyear {2001}}%
}]{%
usher2001time}
\APACinsertmetastar {%
usher2001time}%
\begin{APACrefauthors}%
Usher, M.%
\BCBT {}\ \BBA {} McClelland, J.L.%
\end{APACrefauthors}%
\unskip\
\newblock
\APACrefYearMonthDay{2001}{}{}.
\newblock
{\BBOQ}\APACrefatitle {The time course of perceptual choice: the leaky,
  competing accumulator model.} {The time course of perceptual choice: the
  leaky, competing accumulator model.}{\BBCQ}
\newblock
\APACjournalVolNumPages{Psychological review}{108}{3}{550}.
\newblock

\newblock

\PrintBackRefs{\CurrentBib}

\bibitem [\protect \citeauthoryear {%
Van~Assen%
, Barla%
\BCBL {}\ \BBA {} Fleming%
}{%
Van~Assen%
\ \protect \BOthers {.}}{%
{\protect \APACyear {2018}}%
}]{%
van2018visual}
\APACinsertmetastar {%
van2018visual}%
\begin{APACrefauthors}%
Van~Assen, J.J.R.%
, Barla, P.%
\BCBL {} Fleming, R.W.%
\end{APACrefauthors}%
\unskip\
\newblock
\APACrefYearMonthDay{2018}{}{}.
\newblock
{\BBOQ}\APACrefatitle {Visual features in the perception of liquids} {Visual
  features in the perception of liquids}.{\BBCQ}
\newblock
\APACjournalVolNumPages{Current biology}{28}{3}{452--458}.
\newblock

\newblock

\PrintBackRefs{\CurrentBib}

\bibitem [\protect \citeauthoryear {%
Wang%
, Mei%
, Yuille%
\BCBL {}\ \BBA {} Kortylewski%
}{%
Wang%
\ \protect \BOthers {.}}{%
{\protect \APACyear {2021}}%
}]{%
wang2021neural}
\APACinsertmetastar {%
wang2021neural}%
\begin{APACrefauthors}%
Wang, A.%
, Mei, S.%
, Yuille, A.L.%
\BCBL {} Kortylewski, A.%
\end{APACrefauthors}%
\unskip\
\newblock
\APACrefYearMonthDay{2021}{}{}.
\newblock
{\BBOQ}\APACrefatitle {Neural View Synthesis and Matching for Semi-Supervised
  Few-Shot Learning of 3D Pose} {Neural view synthesis and matching for
  semi-supervised few-shot learning of 3d pose}.{\BBCQ}
\newblock
\APACjournalVolNumPages{Advances in Neural Information Processing
  Systems}{34}{}{}.
\newblock

\newblock

\PrintBackRefs{\CurrentBib}

\bibitem [\protect \citeauthoryear {%
Wong%
, Bi%
, Soltani%
, Yildirim%
\BCBL {}\ \BBA {} Scholl%
}{%
Wong%
\ \protect \BOthers {.}}{%
{\protect \APACyear {2022}}%
}]{%
wong2022seeing}
\APACinsertmetastar {%
wong2022seeing}%
\begin{APACrefauthors}%
Wong, K.W.%
, Bi, W.%
, Soltani, A.A.%
, Yildirim, I.%
\BCBL {} Scholl, B.J.%
\end{APACrefauthors}%
\unskip\
\newblock
\APACrefYearMonthDay{2022}{}{}.
\newblock
{\BBOQ}\APACrefatitle {Seeing soft materials draped over objects: A case study
  of intuitive physics in perception, attention, and memory} {Seeing soft
  materials draped over objects: A case study of intuitive physics in
  perception, attention, and memory}.{\BBCQ}
\newblock
\APACjournalVolNumPages{Psychological Science}{}{}{}.
\newblock

\newblock

\PrintBackRefs{\CurrentBib}

\bibitem [\protect \citeauthoryear {%
Wu%
, Yildirim%
, Lim%
, Freeman%
\BCBL {}\ \BBA {} Tenenbaum%
}{%
Wu%
\ \protect \BOthers {.}}{%
{\protect \APACyear {2015}}%
}]{%
wu2015galileo}
\APACinsertmetastar {%
wu2015galileo}%
\begin{APACrefauthors}%
Wu, J.%
, Yildirim, I.%
, Lim, J.J.%
, Freeman, B.%
\BCBL {} Tenenbaum, J.%
\end{APACrefauthors}%
\unskip\
\newblock
\APACrefYearMonthDay{2015}{}{}.
\newblock
{\BBOQ}\APACrefatitle {Galileo: Perceiving Physical Object Properties by
  Integrating a Physics Engine with Deep Learning} {Galileo: Perceiving
  physical object properties by integrating a physics engine with deep
  learning}.{\BBCQ}
\newblock
 \APACrefbtitle {Advances in Neural Information Processing Systems} {Advances
  in neural information processing systems}\ (\BPGS\ 127--135).
\PrintBackRefs{\CurrentBib}

\bibitem [\protect \citeauthoryear {%
Yamins%
\ \BBA {} DiCarlo%
}{%
Yamins%
\ \BBA {} DiCarlo%
}{%
{\protect \APACyear {2016}}%
}]{%
yamins2016using}
\APACinsertmetastar {%
yamins2016using}%
\begin{APACrefauthors}%
Yamins, D.L.%
\BCBT {}\ \BBA {} DiCarlo, J.J.%
\end{APACrefauthors}%
\unskip\
\newblock
\APACrefYearMonthDay{2016}{}{}.
\newblock
{\BBOQ}\APACrefatitle {Using goal-driven deep learning models to understand
  sensory cortex} {Using goal-driven deep learning models to understand sensory
  cortex}.{\BBCQ}
\newblock
\APACjournalVolNumPages{Nature Neuroscience}{19}{3}{356}.
\newblock

\newblock

\PrintBackRefs{\CurrentBib}

\bibitem [\protect \citeauthoryear {%
Yamins%
\ \protect \BOthers {.}}{%
Yamins%
\ \protect \BOthers {.}}{%
{\protect \APACyear {2014}}%
}]{%
yamins2014performance}
\APACinsertmetastar {%
yamins2014performance}%
\begin{APACrefauthors}%
Yamins, D.L.%
, Hong, H.%
, Cadieu, C.F.%
, Solomon, E.A.%
, Seibert, D.%
\BCBL {} DiCarlo, J.J.%
\end{APACrefauthors}%
\unskip\
\newblock
\APACrefYearMonthDay{2014}{}{}.
\newblock
{\BBOQ}\APACrefatitle {Performance-optimized hierarchical models predict neural
  responses in higher visual cortex} {Performance-optimized hierarchical models
  predict neural responses in higher visual cortex}.{\BBCQ}
\newblock
\APACjournalVolNumPages{Proceedings of the National Academy of
  Sciences}{111}{23}{8619--8624}.
\newblock

\newblock

\PrintBackRefs{\CurrentBib}

\bibitem [\protect \citeauthoryear {%
Yildirim%
, Belledonne%
, Freiwald%
\BCBL {}\ \BBA {} Tenenbaum%
}{%
Yildirim%
\ \protect \BOthers {.}}{%
{\protect \APACyear {2020}}%
}]{%
yildirim2020efficient}
\APACinsertmetastar {%
yildirim2020efficient}%
\begin{APACrefauthors}%
Yildirim, I.%
, Belledonne, M.%
, Freiwald, W.%
\BCBL {} Tenenbaum, J.%
\end{APACrefauthors}%
\unskip\
\newblock
\APACrefYearMonthDay{2020}{}{}.
\newblock
{\BBOQ}\APACrefatitle {Efficient inverse graphics in biological face
  processing} {Efficient inverse graphics in biological face
  processing}.{\BBCQ}
\newblock
\APACjournalVolNumPages{Science Advances}{6}{10}{eaax5979}.
\newblock

\newblock

\PrintBackRefs{\CurrentBib}

\bibitem [\protect \citeauthoryear {%
Yildirim%
, Siegel%
\BCBL {}\ \BBA {} Tenenbaum%
}{%
Yildirim%
\ \protect \BOthers {.}}{%
{\protect \APACyear {2016}}%
}]{%
yildirim2016perceiving}
\APACinsertmetastar {%
yildirim2016perceiving}%
\begin{APACrefauthors}%
Yildirim, I.%
, Siegel, M.H.%
\BCBL {} Tenenbaum, J.B.%
\end{APACrefauthors}%
\unskip\
\newblock
\APACrefYearMonthDay{2016}{}{}.
\newblock
{\BBOQ}\APACrefatitle {Perceiving fully occluded objects via physical
  simulation} {Perceiving fully occluded objects via physical
  simulation}.{\BBCQ}
\newblock
 \APACrefbtitle {Proceedings of the 38th annual conference of the cognitive
  science society.} {Proceedings of the 38th annual conference of the cognitive
  science society.}
\PrintBackRefs{\CurrentBib}

\bibitem [\protect \citeauthoryear {%
Yuille%
\ \BBA {} Kersten%
}{%
Yuille%
\ \BBA {} Kersten%
}{%
{\protect \APACyear {2006}}%
}]{%
yuille2006vision}
\APACinsertmetastar {%
yuille2006vision}%
\begin{APACrefauthors}%
Yuille, A.%
\BCBT {}\ \BBA {} Kersten, D.%
\end{APACrefauthors}%
\unskip\
\newblock
\APACrefYearMonthDay{2006}{}{}.
\newblock
{\BBOQ}\APACrefatitle {Vision as Bayesian inference: analysis by synthesis?}
  {Vision as bayesian inference: analysis by synthesis?}{\BBCQ}
\newblock
\APACjournalVolNumPages{Trends in cognitive sciences}{10}{7}{301--308}.
\newblock

\newblock

\PrintBackRefs{\CurrentBib}

\end{thebibliography}

\bmhead{
  Acknowledgements: This work was supported by the Center for Brains, Minds and Machines (CBMM), funded by NSF STC award CCF-1231216; ONR MURI N00014-13-1-0333 (to J.B.T.); a grant from Toyota Research Institute (to J.B.T.); and a grant from Mitsubishi MELCO (to J.B.T.). A high performance computing cluster (OpenMind) was provided by the McGovern Institute for Brain Research. We thank Kevin Smith, Bernhard Egger, Kelsey Allen, Goker Erdogan, Marty Tenenbaum, Nancy Kanwisher, and Vivian Paulun for their comments on a previous version of this manuscript}

\renewcommand\thefigure{\arabic{figure}}
\setcounter{figure}{0}
\renewcommand{\thetable}{S\arabic{table}}
\renewcommand{\thefigure}{S\arabic{figure}}

\section*{Supplementary Information}

\begin{figure*}[ht!]
    \centering
    \includegraphics[width=0.9\linewidth]{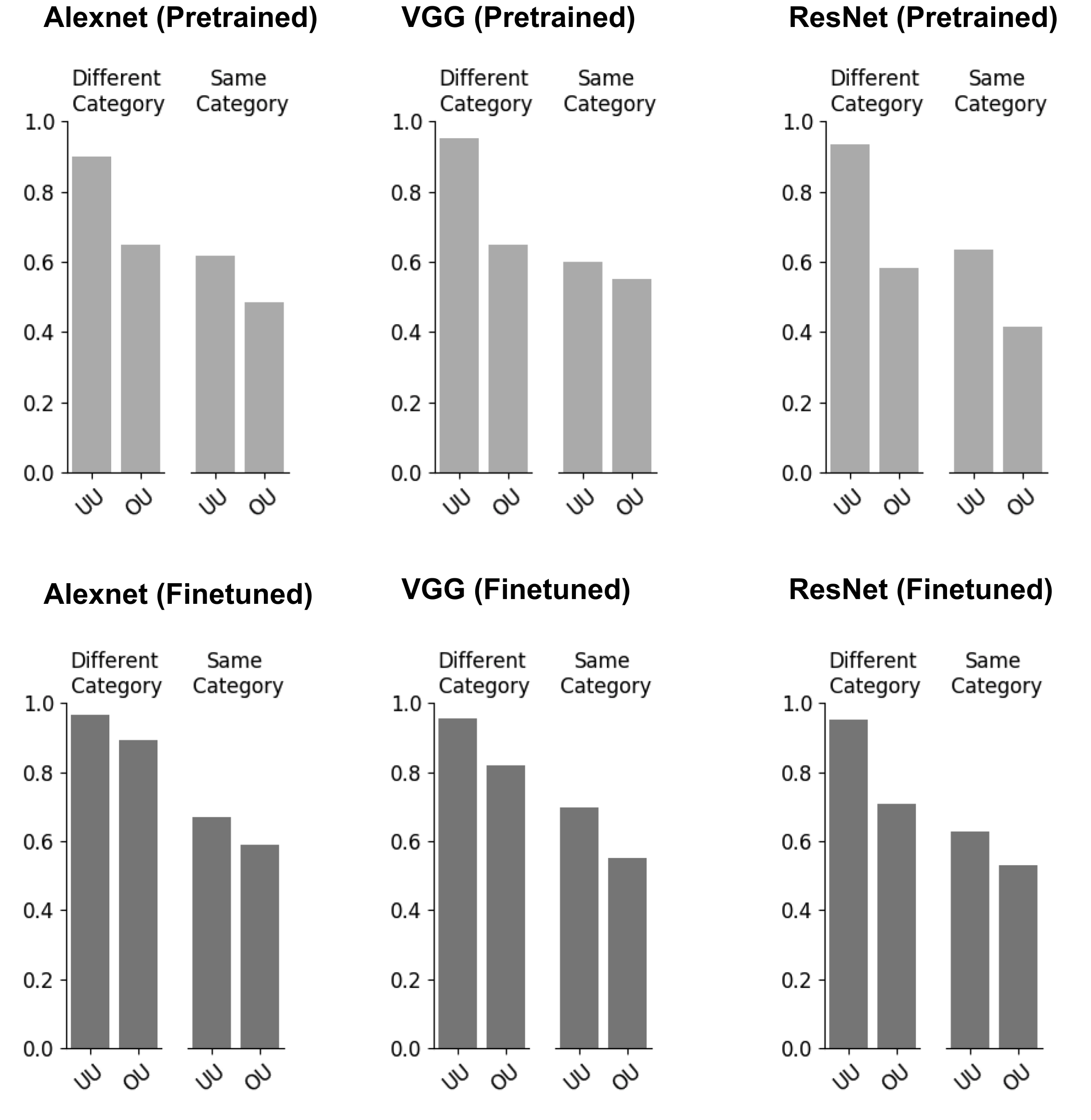}
    \caption{Accuracy levels of the three models we considered including pretrained versions (top row) and after finetuning (bottom row). AlexNet results are presented in the main text.}
    \label{fig:dcnn-results}
\end{figure*}

\begin{figure*}[ht!]
    \centering
    \includegraphics[width=.8\linewidth]{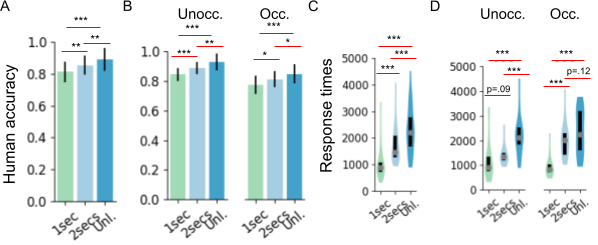}
    \caption{Behavioral results.
    (A) Average human accuracy in the 3 presentation time conditions pooling data across the occlusion conditions. Overall, participants performed well above chance under all presentation time conditions. Behavioral accuracy improved with longer presentation times.
    (B) Average human accuracy shown separately for each occlusion and presentation time condition. Participants' average performance ranged from 73\% in the cloth-occluded condition under 1 sec presentation time to 93\% in the unoccluded condition under unlimited time. The gain in performance was significant within each occlusion condition, $p<.05$ for all pairwise comparisons of presentation times, except in the 1 sec vs. 2 secs comparison in the cloth-occluded condition, $p=.07$. %
    (C) Average response times (in milliseconds; pooling data across the occlusion conditions) lengthen with longer presentation times, $p<.001$ for all pairwise comparisons of presentation time conditions.
    (D) Average response times shown separately for each occlusion and presentation time condition. Lengthening of response times is still evident for each occlusion condition ($p<.05$ for all pairwise comparisons of presentation times, except in the 1 sec vs. 2 secs comparison in the unoccluded condition and 1 secs vs. Unl. comparisons in the cloth-occluded condition). %
    Error bars show standard deviation.}
    \label{fig:behavior-results}
\end{figure*}

\begin{figure*}[ht!]
    \centering
    \includegraphics[width=.8\linewidth]{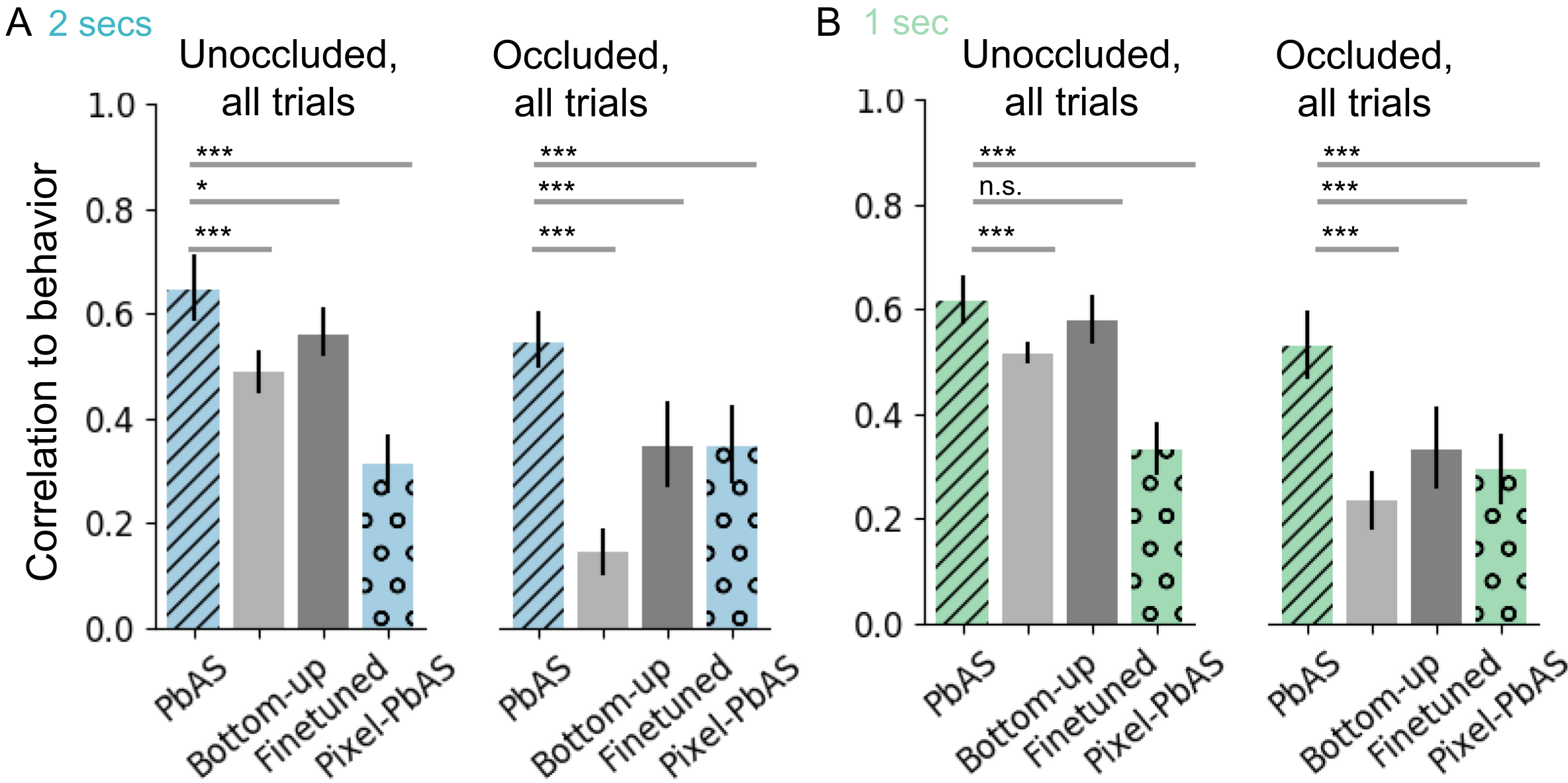}
    \caption{Trial-level accuracy correlations in the (A) 2 secs presentation time condition and (B) 1 sec presentation time conditions. The physics-based analysis-by-synthesis (PbAS)  model correlates well with behavior across all presentation time and occlusion condition time conditions, relative to the alternatives based on bottom-up features optimized for image classification (BU: bottom-up network with pretrained weights from ImageNet dataset; FT: fine-tuned networks, separately fine-tuned for each occlusion conditions) and Pixel-PbAS, an ablation of PbAS without the bottom-up image encoding modules (using pixels directly for likelihood computation). Significance convention same as main text. Error bars indicate bootstrapped 95\% confidence intervals.}
    \label{fig:supp-correlations}
\end{figure*}

\begin{figure*}[ht!]
\centering
\includegraphics[width=.8\linewidth]{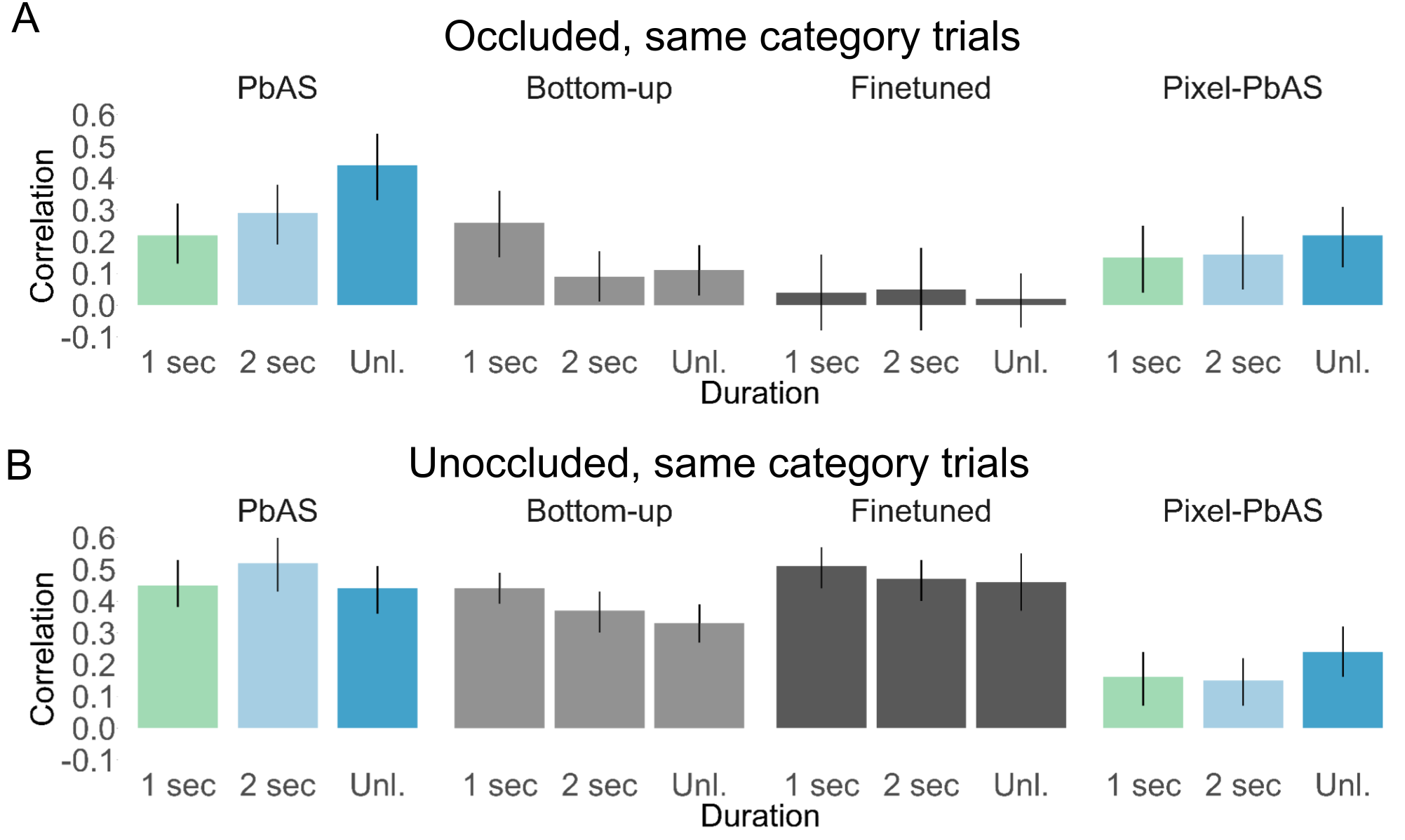}
\caption{Trial-level accuracy correlations for the difficult, same category trials in the (A) Occluded and (B) Unoccluded conditions. Results are arranged by model type and stimulus presentation time. Error bars show bootstrapped 95\% confidence intervals. %
In the easier unoccluded, shape-category conditions, all three models that use DCNN features to match images (PbAS, BU, and FT) perform similarly across all presentation times; Pixel-PbAS performs significantly worse across all presentation times.  In the more difficult occluded, same-category conditions, PbAS clearly outperforms all other models, except for BU which performs similarly in the shortest (1 sec) presentation time.
Notably both pure DCNN models, BU and FT, consistently correlate less well with human trial-level accuracies as presentation times increase, while PbAS correlations tend to increase, and FT correlations are not significantly different from zero in the challenging occluded same-category conditions (with BU correlations being only barely higher than zero in the 2 sec and unlimited conditions).  This overall pattern is consistent with the success of DCNNs at capturing the rapid feedforward contributions to human object recognition for familiar stimuli viewed under standard conditions, and strengthens our proposal that more challenging viewing conditions and longer processing times engage top-down, iterative, generative model based computations of the form instantiated in PbAS.  The combination of physics-based analysis by synthesis with DCNN features for matching generative model simulations to images, as instantiated in the full PbAS model but not Pixel-PbAS, is the only model that accounts well (and better than or equal to any other model) for all stimulus conditions and all presentation times.
}
\label{fig:supp-correlations-same-category}
\end{figure*}

\begin{figure*}[ht!]
    \centering
    \includegraphics[width=.7\linewidth]{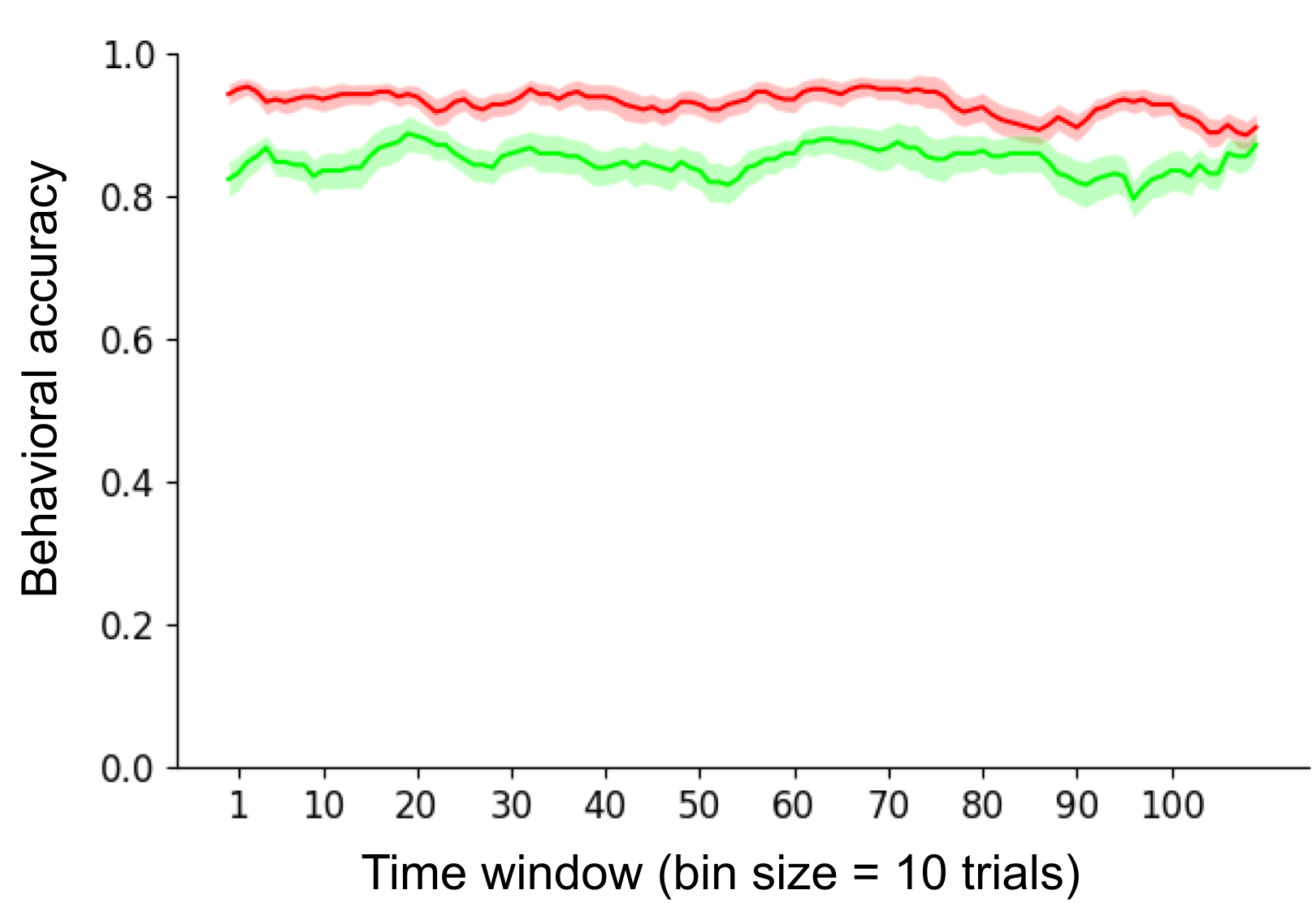}
    \caption{Behavioral learning curves in the two occlusion conditions. We show moving window averages (window size=10) of human accuracy levels in the two occlusion conditions (red=UU, green=OU) under the unlimited presentation time condition. We find no evidence of learning throughout the experiment. Shaded region shows standard error.}
    \label{fig:learning-curves}
\end{figure*}

\begin{figure*}[ht!]
    \centering
    \includegraphics[width=.7\linewidth]{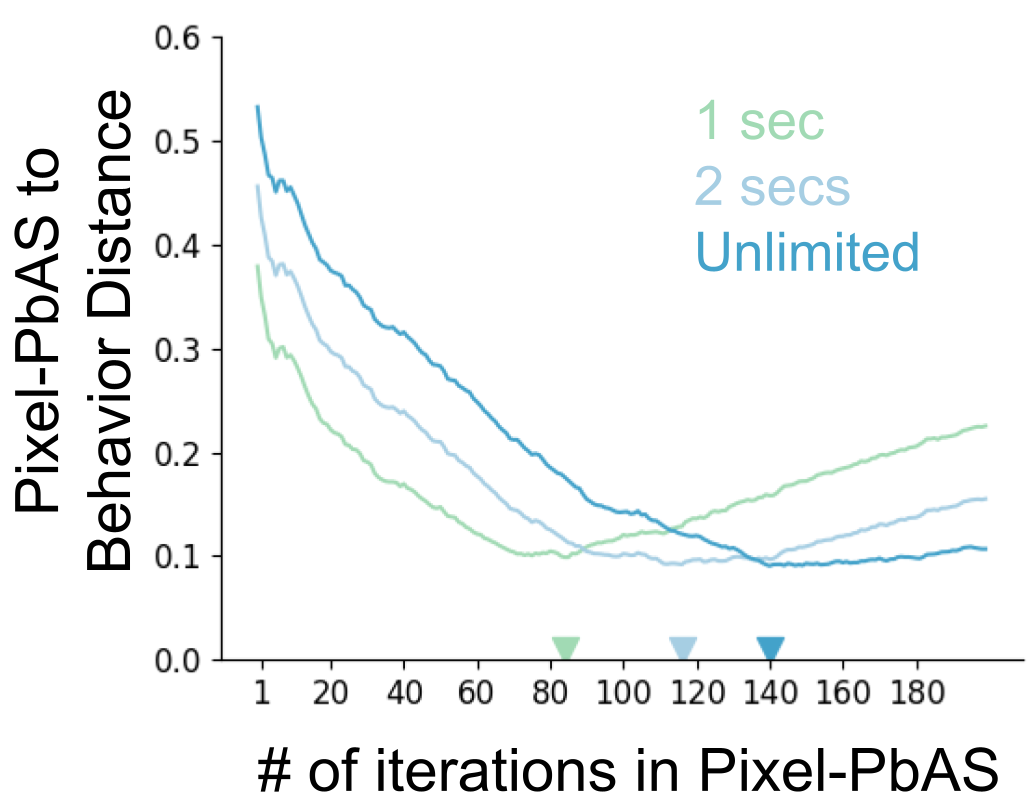}
    \caption{Divergence between the Pixel-PbAS model (i.e., using pixels for likelihood without bottom-up image encoding) and human performance at each model iteration. Colored lines show $\ell_2$ distance between this model and human accuracy for all trials in indicated presentation time condition. Colored triangles indicate the best matching iterations for each presentation time condition. This model asymptotes at a larger distance to behavior than the PbAS model.}
    \label{fig:alternative-losses}
\end{figure*}

\end{document}